\PassOptionsToPackage{colorlinks=true,bookmarksopen=true,bookmarksnumbered=true,citecolor=blue,urlcolor=blue,linkcolor=blue}{hyperref}
\documentclass[12pt,letterpaper]{article}

\usepackage[T1]{fontenc}
\usepackage[a4paper, total={7in, 10in}]{geometry}

\usepackage{graphicx}
\usepackage{helvet}
\usepackage{authblk}
\usepackage{amsmath}
\usepackage{amssymb}
\usepackage{orcidlink}
\usepackage[super,comma,sort&compress]{natbib}
\usepackage[right]{lineno}

\usepackage{booktabs}
\usepackage{caption}
\usepackage{xcolor}
\usepackage[many]{tcolorbox}
\usepackage{varwidth}
\usepackage{enumitem}
\usepackage{tabularx}

\usepackage[frozencache,cachedir=minted]{minted}
\usepackage[scaled=0.85]{beramono}
\definecolor{bg}{rgb}{0.97,0.97,0.97}
\setminted{fontsize=\small,bgcolor=bg}

\usepackage{hyperref}

\linenumbers

\definecolor{task}{RGB}{222,235,247}
\definecolor{context}{RGB}{253,219,199}
\definecolor{input}{RGB}{213,232,212}
\definecolor{output}{RGB}{249,237,187}

\tcbset{
  enhanced,
  sharpish corners,
  before skip=10pt, after skip=10pt,
  boxsep=1em, left=1em, right=1em, top=0.8em, bottom=0.8em
}

\newtcolorbox{headerbox}[2][]{%
  colback=white,
  coltext=black,
  boxrule=0pt,
  frame hidden,
  fuzzy shadow={0pt}{-2pt}{-0.5pt}{0.5pt}{black!30},
  title={#2},
  fonttitle=\bfseries\small,
  colbacktitle=black!85,
  coltitle=white,
  toprule=0pt,
  boxed title style={sharpish corners, boxrule=0pt, interior style={fill=black!85}},
  attach boxed title to top center={yshift*=-\tcboxedtitleheight/2},
  varwidth boxed title*=\linewidth,
  before skip=2em,
  after skip=2em,
  borderline north={0.6pt}{0pt}{black!75},
  #1
}

\setlist[itemize]{topsep=2pt plus 2pt minus 2pt,itemsep=2pt,parsep=0pt,leftmargin=1.2em}
\setlist[enumerate]{topsep=2pt plus 2pt minus 2pt,itemsep=2pt,parsep=0pt,leftmargin=1.2em}

\makeatletter
\renewcommand{\maketitle}{\bgroup\setlength{\parindent}{0pt}
\begin{flushleft}
  \textbf{\@title}

  \@author
\end{flushleft}\egroup}
\makeatother

\title{A Methodological Guide on Using Large Language Models for Reproducible Text Annotation in the Social Sciences and Humanities with Python and R}
\date{}

\author[1,2]{Qixiang Fang}
\author[1]{Javier Garcia Bernardo}
\author[1]{Erik-Jan van Kesteren}

\affil[1]{ODISSEI Social Data Science Team, Utrecht University, Utrecht, The Netherlands}
\affil[2]{Lead contact}
\affil[*]{Correspondence: q.fang@uu.nl}

\begin{document}

\maketitle

\section*{SUMMARY}
Large language models (LLMs) are increasingly used by researchers in the social sciences and humanities (SSH) for text analysis, particularly to automate text annotation. However, many researchers still face challenges in adopting LLMs, addressing their limitations, and producing reproducible workflows and results. For example, annotation errors can bias downstream statistical analyses even when apparent accuracy is high. This paper provides a step-by-step methodological guide to using LLMs for text annotation in SSH research, with practical Python and R examples. We explain how LLMs work, how to set up research projects, how to interact with (open-source) LLMs programmatically, how to design and evaluate prompts without overfitting, how to integrate LLM annotations into statistical analyses while accounting for annotation error, and how to manage cost, efficiency, and reproducibility at scale. Throughout, we emphasize intuitive methodological reasoning, concrete examples, and best practices to help researchers incorporate LLM-based annotation into reproducible scientific workflows.

\section*{KEYWORDS}

large language models, text annotation, social sciences, humanities, measurement, regression, Python, R, annotation error, reproducibility, local deployment, API, metadata

\section*{INTRODUCTION}

\setcounter{secnumdepth}{3}
\begin{figure}[ht]
    \centering
    \includegraphics[width=\linewidth]{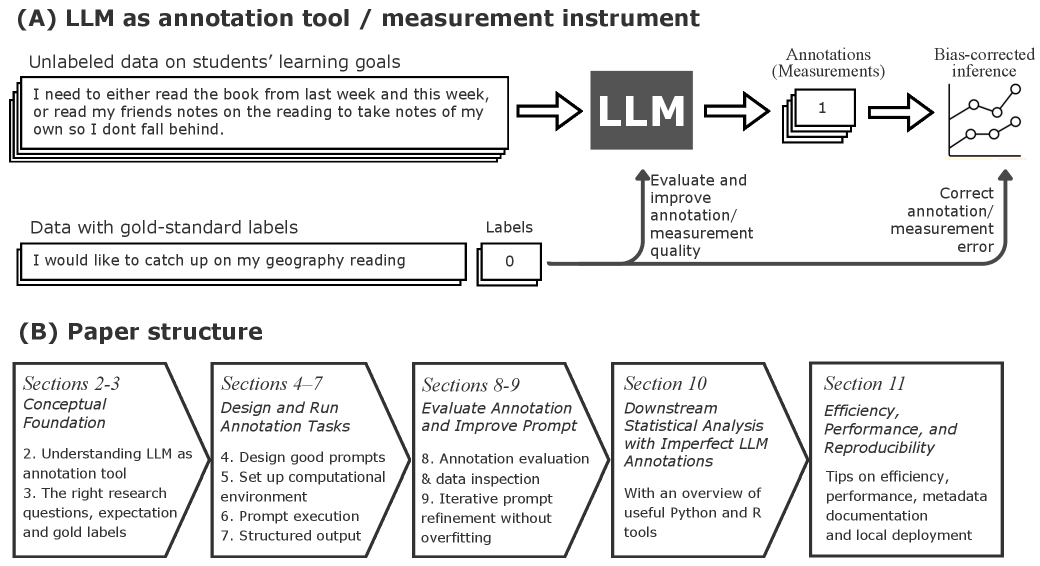}
    \caption{Conceptual and structural overview of this paper.}
    \label{fig:overview}
\end{figure}

\section{Introduction}
\label{sec:introduction}
The recent proliferation of large language models (LLMs), such as GPT-5, Qwen and Llama, has transformed how researchers annotate and analyze text data~\citep{brown2020language, openai2024gpt4}. Text annotation is a fundamental component of empirical research in the social sciences and humanities (SSH), serving as the basis for content analysis, sentiment measurement, stance detection, and narrative framing studies across many domains~\citep[e.g.,][]{krippendorff2018, grimmer2013text, mohammad2016sentiment,ziems2024css}. However, large-scale manual annotation is often costly, slow, and inconsistent across annotators~\citep{artstein2008inter,poletto2021resources,li2023coannotating}.

LLMs offer a promising way to automate or augment this process~\citep{gilardi2023chatgpt, ziems2024css}. Their central promise is that text annotation tasks can be performed automatically, cheaply, and correctly with a few well-crafted prompts instead of extensive manual labelling.

Yet for many SSH researchers, barriers to adopting LLMs in their research workflows remain substantial. \emph{First}, understanding how LLMs work and what their limitations imply for research is not straightforward, especially for researchers without relevant training backgrounds. Selecting suitable models, interacting with LLMs programmatically, conducting rigorous performance evaluation, and keeping up with rapid model development can be challenging and overwhelming for newcomers. \emph{Second}, even experienced researchers often underestimate the impact of annotation error on subsequent analyses. Imperfect labels, no matter how high the apparent annotation accuracy, can propagate into downstream statistical analyses, biasing regression coefficients, inflating or deflating \textit{p}-values, and ultimately affecting substantive scientific conclusions~\citep{egami2023dsl}. \emph{Third}, LLM-based workflows raise distinctive reproducibility challenges. For example, proprietary cloud models may change silently over time while older versions may no longer remain publicly accessible. Also, regardless of whether LLMs are proprietary or open-source, seemingly minor differences in prompts, model settings, and even computational environments can affect annotation outputs.

This paper seeks to address these challenges. While several recent reviews have surveyed how LLMs and broader NLP methods are used in SSH research~\citep[e.g.,][]{ziems2024css,Abdurahman2025}, our goal is distinct:
we provide a \textit{methodologically driven, code- and workflow-focused guide} that enables researchers to quickly implement, evaluate, and improve their own LLM-based annotation workflows. With an emphasis on transparent, reproducible, and research-centered practice, we aim to help both novice and experienced readers incorporate LLMs into their projects with ease, confidence and scientific rigor.

Concretely, this paper \textbf{is} about:
(1) a concise, SSH-centered overview of how LLMs work \emph{for text annotation purposes} and what they can (and cannot) do;
(2) a scientific and reproducible API-based workflow for text annotation for both Python and R users, with a focus on open-source and local LLMs;
(3) methods to evaluate LLM annotation quality, identify issues, and iteratively improve performance without overfitting;
(4) techniques for incorporating annotation uncertainty in downstream statistical analyses; and
(5) further deployment and reproducibility tips for scaling LLM annotation projects.

This paper \textbf{is not} about:
(1) exact technical or architectural details of LLMs; (2) benchmarking or ranking LLMs for SSH applications; (3) fine-tuning LLMs for domain adaptation; or (4) \emph{replacing the need for gold-standard human annotations entirely}. Specifically, we situate LLMs as a new class of annotation tools or, more broadly, as \textit{measurement instruments}: powerful yet imperfect tools that can augment (not replace) human expertise.

The remainder of this paper is organized as follows.

\begin{itemize}
\item \hyperref[sec:llm]{Section~\ref*{sec:llm}: \nameref*{sec:llm}} provides an introduction to LLMs through the lens of logistic regression, their use for text annotation, and their limitations.

\item \hyperref[sec:projectsetup]{Section~\ref*{sec:projectsetup}: \nameref*{sec:projectsetup}} discusses how to set up a research project, with a focus on suitable research questions and data with ground-truth labels.

\item \hyperref[sec:prompt]{Section~\ref*{sec:prompt}: \nameref*{sec:prompt}} focuses on designing good prompts, often the most consequential decision when using LLMs for text annotation.

\item \hyperref[sec:env]{Section~\ref*{sec:env}: \nameref*{sec:env}} describes how to use cloud notebooks to quickly build a lightweight computational environment for API-based LLM annotation.

\item \hyperref[sec:initial]{Section~\ref*{sec:initial}: \nameref*{sec:initial}} demonstrates how to submit a first prompt to an LLM API and obtain an LLM response.

\item \hyperref[sec:struct]{Section~\ref*{sec:struct}: \nameref*{sec:struct}} shows how to structure LLM responses to improve both annotation quality and efficiency.

\item \hyperref[sec:evaluation]{Section~\ref*{sec:evaluation}: \nameref*{sec:evaluation}} discusses how to evaluate annotation quality, how to diagnose data and prompt issues, and how to iterately improve performance.

\item \hyperref[sec:best_prompt]{Section~\ref*{sec:best_prompt}: \nameref*{sec:best_prompt}} extends \autoref{sec:evaluation} and proposes a three-step framework (Explore $\rightarrow$ Select $\rightarrow$ Evaluate) to help researchers find the best prompt while accounting for the issues of overfitting and biases in performance estimation.

\item \hyperref[sec:downstream]{Section~\ref*{sec:downstream}: \nameref*{sec:downstream}} presents types of annotation error and methods to account for them in downstream statistical analyses.

\item \hyperref[sec:nextlevel]{Section~\ref*{sec:nextlevel}: \nameref*{sec:nextlevel}} offers additional tips on performance, efficiency, and reproducibility, including metadata documentation and local model deployment, which are especially important when scaling up a project.

\end{itemize}

Together, these sections bridge the gap between conceptual understanding of LLMs and technical implementation, enabling SSH researchers to use LLMs critically, scientifically, and effectively for text annotation in an open-scientific, reproducible manner. Finally, the paper is accompanied by two Python and two R codebooks that mirror the steps in the text.\footnote{Access the codebooks via \url{https://github.com/sodascience/workshop_llm_data_collection} or \url{https://zenodo.org/records/20073016}. Additional learning materials including slides and resource guides are available.}

\section{Understand LLMs for Text Annotation}
\label{sec:llm}
LLMs such as Llama, Qwen, Claude, and ChatGPT are built on a family of deep learning architectures known as \textit{transformers}~\citep{vaswani2017attention}.
At their core, they are predictive models: given a sequence of text, they estimate the probability (distribution) of what comes next in the form of \textit{tokens}. Tokens are the smallest units recognized by the model, including words, subwords, and punctuation.
Although the underlying mathematics involves billions of parameters and matrix computations, the core logic is familiar to many SSH researchers with experience in logistic regression.

\subsection{From Logistic Regression to Language Modelling}
In logistic regression, we model the probability of an outcome $y \in \{0,1\}$ (e.g., ``negative'' vs. ``positive'') as a logistic transformation of a weighted combination of predictors:
\[
P(y=1 \mid \mathbf{x}) = \frac{1}{1 + \exp(-\boldsymbol{\beta}^{\top}\mathbf{x})}
\]
where $\mathbf{x}$ is a vector of predictor values $(x_1, x_2, \ldots, x_p)$ and $\boldsymbol{\beta}$ is a corresponding vector of coefficients $(\beta_1, \beta_2, \ldots, \beta_p)$.
Each coefficient $\beta_j$ captures how strongly predictor $x_j$ influences the log-odds of the outcome.

An LLM performs a conceptually similar operation, but at a much larger scale.
Instead of a handful of (typically manually specified) predictors, it internally computes thousands of \textit{latent features} that represent meaningful patterns in the input text.
These latent features are then combined and weighted, much like predictors in logistic regression, and transformed into a probability distribution over all possible next tokens in the model's vocabulary.
Because the outcome space consists of many possible tokens rather than \textit{two} classes, a \textit{softmax} transformation is used instead of the logistic function:
\[
P(\text{next token} = t_i \mid \text{input text}) =
\frac{\exp(\mathbf{w}_i^{\top}\mathbf{h})}{\sum_j \exp(\mathbf{w}_j^{\top}\mathbf{h})},
\]
where the latent feature vector $\mathbf{h}$ represents meaningful patterns in the input text (analogous to predictor vector $\mathbf{x}$ in logistic regression), and $\mathbf{w}_i$ is a learned weight vector for each possible next token $t_i$ (analogous to coefficient vector $\boldsymbol{\beta}$ in logistic regression).
Both $\mathbf{h}$ and $\mathbf{w}_i$ share the same dimensionality (i.e., length), and the denominator sums over all tokens in the vocabulary.
Conceptually, this makes an LLM a massively multivariate extension of logistic regression operating over text sequences.

\subsection{A Segue to Prompting and Annotation}
When provided with a simple prompt such as \textit{Classify the student's goal as specific or vague:''}, an LLM conditions on this input and generates the most probable continuation (e.g., ``specific’’ or ``vague’’). This illustrates how prompting can cast a task as a constrained text completion problem, in which the desired outputs correspond to a predefined set of labels. From this perspective, LLMs can be viewed as flexible text classifiers~\citep{gilardi2023chatgpt, ziems2024css}, supporting a range of annotation tasks in SSH research, such as classifying text into theoretically meaningful categories and extracting entities or themes from open-ended responses.

However, several caveats are worth noting:
\begin{itemize}
    \item \textbf{Probabilistic nature.} LLMs produce the most likely continuation, not necessarily the \emph{true} or \emph{correct} label. 
    \item \textbf{Reproducibility issues.} Proprietary cloud-based models may change silently over time. Recording the model version and prompt text can improve reproducibility, but recovering older models and outputs can still be difficult. Furthermore, regardless of whether LLMs are open-source or proprietary, small variations in model settings, computational environments, or prompt phrasing can yield slightly, or even substantially, different outputs.
    \item \textbf{Opacity of features.} Unlike logistic regression, where $\beta$ coefficients have interpretable meanings, the internal weights of LLMs are distributed across billions of parameters and cannot be directly interpreted~\citep{bender2021dangers}.
    \item \textbf{Training data and developer dependence.} Model behavior reflects the distribution of its training data and the choices made by its developers, which may include biases or outdated information~\citep{mitchell2019model}.
    \item \textbf{Hidden costs.} Beyond API usage fees, employing LLMs for annotation requires time and resources for researcher training, prompt engineering, and iterative experimentation. For smaller projects, manual annotation can still be cheaper and more efficient.
\end{itemize}

\subsection{Summary}
For SSH researchers, it is often useful to think of an LLM as an ``expanded logistic regression'' trained on large-scale internet text: it repeatedly predicts the next token/label using internally computed latent features that represent patterns in the input.
When properly instructed through prompts, this predictive machinery can be repurposed for many SSH annotation tasks.

\begin{headerbox}{Running Example: Goal Setting and Planning in Learning}
Throughout the remainder of this paper, we use a running example from educational research: annotating weekly dialogues between a student and a \emph{rule-based chatbot}, with the goal of evaluating the \emph{specificity of the student's goal setting and planning}. The chatbot asks students a \emph{fixed} list of questions designed to help them set better learning goals. It functions as a structured questionnaire presented through a user-friendly conversational interface, rather than as an intelligent assistant such as an LLM.
The accompanying dataset was kindly provided by Gabrielle Martins van Jaarsveld as part of her research on self-regulated learning~\citep{VanJaarsveld2025}. A dialogue in the dataset looks as follows:

\begin{quote}
\small
\textbf{Chatbot:} Set an academic goal for the upcoming week.\\
\textbf{Student:} I would like to catch up on my geography reading\\
\textbf{Chatbot:} Add details to make your goal more specific.\\
\textbf{Student:} I need to either read the book from last week and this week, or read my friends notes on the reading to take notes of my own so I dont fall behind.\\
\textbf{Chatbot:} How will you measure progress on and achievement of your goal?\\
\textbf{Student:} by the number of pages I write per day\\
\textbf{Chatbot:} Why is this goal important to you in the context of your prior experiences and future goals?\\
\textbf{Student:} It is important to achieve because if I dont, I will fall behind and most likely wont be ready for the exam.\\
\textbf{Chatbot:} Create a step-by-step plan for achieving this goal in the coming week.\\
\textbf{Student:} 1. evaluate how much there is to do\\
2. get help from my friends\\
3. takes notes day by day
\end{quote}
\end{headerbox}

\section{Set Up The Research Project}
\label{sec:projectsetup}
In this section, we discuss how to set up a research project in a way that aligns with the requirements of LLM-based text annotation.

\subsection{Know Your Research Question and Data}
Begin by articulating your research questions and evaluating whether your data can meaningfully address them. In other words, ensure that what you aim to measure is actually present in the textual materials available to you. Some SSH constructs, such as personality traits, are inherently difficult to capture from data sources like social media posts because of limited linguistic and contextual cues~\citep{stajner_why_2021,fang2023text}. Similarly, in highly subjective annotation tasks such as hate speech detection~\citep{leonardelli2021agreeing}, human annotators will inevitably disagree. In both cases, LLMs are likely to encounter similar challenges.

It is therefore important to recognize that LLMs are not universally suitable for all annotation tasks. When a construct is weakly encoded in text, poorly defined, or highly dependent on contextual knowledge that is unavailable in the dataset, even state-of-the-art models will struggle. LLMs do not magically resolve conceptual ambiguity or limited signal; they reproduce and sometimes amplify the constraints inherent in the data and task design.

Researchers should therefore set realistic expectations for LLM performance. Performance ceilings are often determined not by model capacity, but by construct clarity, annotation guidelines, inter-annotator (dis)agreement, and the information density of the text. If human annotators reach only moderate agreement, expecting near-perfect alignment from an LLM is neither realistic nor methodologically justified.

If it is deemed appropriate to proceed with LLM-based annotation, careful planning becomes essential. Before beginning annotation, researchers should define an \emph{a priori} target level of agreement between the LLM and gold-standard labels that would render the LLM-generated annotations fit for downstream use. Crucially, this target should be theoretically and substantively motivated (e.g., sufficient for reliable group comparisons, predictive modeling, or statistical significance testing), rather than driven by arbitrary benchmarks or post hoc justifications. Establishing such a target in advance clarifies what constitutes ``good enough’’ performance and provides a principled stopping criterion for prompt refinement or model iteration. In doing so, it also helps prevent excessive optimization on a validation set, which may inflate apparent performance (i.e., overfitting) without improving generalizability.

In short, strong LLM-based annotation begins not with the model, but with appropriate research questions, clear construct and concept definitions, strong data, and defensible performance criteria.

\begin{headerbox}{Running Example: Agreement and Expectation}
In a pilot study involving four human annotators with PhD-level expertise, inter-annotator agreement exceeded 0.85.
Given this high agreement score, the clarity of the rubric, and the presence of explicit textual cues in the text, we can expect similar performance from an LLM on this task.
\end{headerbox}

\subsection{Secure a Gold-Labelled Subset}
Establishing \emph{gold-standard labels} (also called \emph{gold labels} or \emph{ground-truth labels}) for a subset of your data is essential for rigorously evaluating the performance of LLM-generated annotations. These labels typically come from a carefully curated subset of texts annotated using well-defined guidelines, ideally by domain experts or objective criteria, and serve as the highest-quality reference against which model outputs can be compared.

Aim for balance across categorical labels or ranges of numeric scores so that evaluation is informative rather than dominated by a few categories or score ranges. If you anticipate subgroup analyses (e.g., by time period or data source), stratify the gold-standard subset accordingly to ensure that performance estimates reflect those intended use cases.

Crucially, keep this gold-standard subset \emph{strictly held out} from prompt examples, model fine-tuning, and training data to prevent data leakage~\citep{kapoor2023leakage}. Leakage can artificially inflate agreement between LLM-generated and gold-standard labels and lead to overly optimistic conclusions about model performance.

Obtaining such a subset usually requires human annotation. In typical SSH research, several scenarios are possible:

\begin{itemize}
    \item You already have a dataset with a subset of gold labels. These labels may be based on objective information (e.g., authorship, book genre) or expert annotations (often by following a validated codebook or rubric).
    \item You have a validated codebook or rubric and only need to recruit new annotators to produce gold-standard labels.
    \item You have neither of the above and need to develop a new codebook and collect gold-standard annotations from scratch.
    \item You plan to use a strong (and ideally different) LLM to generate an initial gold-labelled subset, followed by manual verification. This mixed approach is intended to save time and resources.
\end{itemize}

\begin{headerbox}{Running Example: Rubric for Goal Specificity}
Below is the rubric that Gabrielle Martins van Jaarsveld developed by synthesizing prior research and expert input:

\begin{quote}
\small
\begin{itemize}
    \item \textbf{Characteristic Measured:} Goal specificity
    \item \textbf{Definition:} Goal must be specific rather than general. The context and details of the goal should be explicitly described, and all abstract terms are explained.
    \item \textbf{Scoring Rules:}
    \begin{itemize}
        \item \textbf{Not Present (0):} Extremely broad, with no details about what this goal entails. States the goal using vague terms without providing any descriptions of what they mean. \textit{Example student response: ``Prepare for the tutorial.''}
        \item \textbf{Partially Present (1):} States the goal and offers some descriptions of the terms used, however there are still some vague terms which are not fully described. \textit{Example student response: ``Prepare for this week's Friday morning tutorial by reading all the materials.''}
        \item \textbf{Fully Present (2):} No vague terms which are not described. Clearly states the goal and uses clear descriptions to describe exactly what they want to achieve. OR Gives a boundary descriptor which offers context to the other ‘unexplained' terms in the goal. \textit{Example student response: ``Prepare for this week's Friday morning tutorial by reading all 5 of the articles which are assigned on canvas.''}
    \end{itemize}
    
\end{itemize}
\end{quote}
\end{headerbox}

Regardless of the scenario, it is crucial to ensure a strong conceptual link between your research question and the annotation instructions. The codebook or rubric serves as this bridge: it operationalizes theoretical constructs and later forms the backbone of prompt design, which is arguably the most critical step in LLM-based text annotation.

A strong codebook or rubric is \emph{clear}, \emph{complete}, and \emph{easy to follow}, while avoiding unnecessary jargon and ambiguity. Drawing on prior research, established rubrics, and domain expertise can substantially improve quality. Although a detailed discussion of best practices for obtaining high-quality human annotations is beyond the scope of this paper~\citep[see, e.g.,][]{wynne2005developing}, it is important to recognize that LLM-based and human annotation are conceptually similar processes: \emph{both depend heavily on precise, well-structured, and unambiguous instructions}.

\paragraph{Caveat}
Even carefully constructed gold-labelled subsets may contain errors or inconsistencies, which can create apparent disagreement between LLM-generated and gold-standard labels. In \autoref{sec:downstream}, we discuss strategies for diagnosing and accounting for such discrepancies.

\subsection{Summary}
Before writing prompts or code, ensure that: (1) the data contain the signals required to address your annotation tasks and research questions; (2) a realistic target agreement threshold has been defined; (3) your research questions have been translated into clear annotation instructions (e.g., a codebook, rubric, or scoring manual); and (4) a balanced, held-out gold-standard subset is available for evaluation.

\section{Design Good Prompts}
\label{sec:prompt}

\subsection{Taxonomy and Anatomy of Prompts}
A prompt is defined as natural language input or instructions (i.e., text) that describe the task that an LLM (and other AI systems) should perform~\citep{brown2020language}. 
Prompt design is the most crucial part of obtaining consistent and interpretable annotations from LLMs. 

Most LLM systems separate a \emph{system prompt} (which sets overall model behavior) and a \emph{user prompt} (which contains the task input). 
In the context of LLM-based annotations, 
system prompts should describe the task \colorbox{context}{context}, the concrete \colorbox{task}{task} itself, and the expected \colorbox{output}{output}. 
In contrast, user prompts provide the \colorbox{input}{input} to the task (or say task content).

\begin{headerbox}{Running Example: Basic Prompt Design}
\textbf{System Prompt: }

\colorbox{context}{A university student was given a series of questions from a chatbot, guiding} \colorbox{context}{them through the
process of setting and elaborating on an academic goal for} \colorbox{context}{the coming week.}\colorbox{task}{You
will be provided with the entire conversation including} \colorbox{task}{the chatbot's questions and the student's answers. Your objective is to assess} 
\colorbox{task}{the specificity of the student's goal} \colorbox{output}{on a scale of 0-2} \colorbox{task}{based on the entire conversation.} 

\vspace{1em}
\textbf{User Prompt: }

\colorbox{input}{\textbf{Chatbot:} Set an academic goal for the upcoming week.}
\\
\colorbox{input}{\textbf{Student:} I would like to catch up on my geography reading}\\
\colorbox{input}{\textbf{Chatbot:} Add details to make your goal more specific.}\\
\colorbox{input}{[The rest of the conversation is omitted.]} 
\end{headerbox}

\subsection{Prompt Engineering Techniques (Quick Wins)}
Prompt engineering refers to the process of modifying instructions to produce better outputs from LLMs.
Several simple adjustments can markedly improve annotation performance.

\begin{enumerate}
  \item \textbf{Clarity.} Be explicit. Avoid ambiguity and vague wording. For instance, verbs and phrases such as ``classify,'' ``assign one label,'' and ``choose the best option'' are preferred over ``analyse'' and ``interpret''. Avoid common pitfalls such as vague structure, missing contextual information, or inconsistent terminology. If a prompt appears ambiguous to humans, it will almost certainly confuse the LLM~\citep{openai2024gpt4}.
  \item \textbf{Role-based prompting.} Assign a concise role (e.g., ``You are an educational expert'') to guide the LLM's response style, tone, and task interpretation~\citep{openai2024gpt4}.
  \item \textbf{Step-by-step reasoning.} When the task is complex, instruct the model to reason step by step (internally) before giving the final label. This approach is also known as chain-of-thought prompting~\citep{wei2022chain}.
  \item \textbf{Few-shot prompting.} Provide a few labelled examples within the system prompt to anchor decisions. This is especially helpful for nuanced categories~\citep{brown2020language}.
\end{enumerate}

Below, we show how to incorporate these four techniques into the system prompt using our running example.

\begin{headerbox}{Running Example: Improving Clarity in System Prompt}
The following task description provides substantially greater clarity regarding the concept of goal specificity and score interpretation. 

\begin{quote}
\small
\#\#TASK\#\#

A university student was given a series of cues by a chatbot, guiding them through the process of setting and elaborating on an academic goal for the coming week. You will be provided with the entire conversation including the chatbot cues, and the student answers. Your objective is to assess all the students answers using a scoring rubric that evaluates the specificity of the goal on a scale of 0 to 2, representing the characteristic being not present, partially present, or fully present. 
Specificity is defined as: Goal must be specific rather than general. The context and details of the goal should be explicitly stated and described, and all terms are explained.

- Score of 0: Extremely broad, with no details about what this goal entails. States the goal using vague terms without providing any descriptions of what they mean. Or the goal is an abstract concept to improve or work towards, without any explanation of how this could be actionable or concrete.

- Score of 1: States an actionable or concrete goal and offers some descriptions of the terms used. However, there are still some vague terms which are not fully described.

- Score of 2: No vague terms which are not described. Clearly states the goal and uses clear descriptions to describe exactly what they want to achieve. OR gives a boundary descriptor which offers context to the other unexplained terms in the. 

\end{quote}
\end{headerbox}

\begin{headerbox}{Running Example: Using Role-based Prompting in System Prompt}
Insert the following role description at the beginning of your system prompt: 

\begin{quote}
\small
You are an expert in educational assessment and goal evaluation, with specialized expertise in applying deductive coding schemes to score the quality and content of student goals. You have a deep understanding of scoring rubrics and are highly skilled at analysing goals for specific characteristics according to well-defined criteria.
\end{quote}

\end{headerbox}

\begin{headerbox}{Running Example: Integrating Step-by-Step Reasoning in System Prompt}
The text below provides clear, step-by-step reasoning instructions.

\begin{quote}
\small

\#\#INSTRUCTIONS\#\#

1. Understand the scoring rubric:
REVIEW the rubric provided for specificity to understand the criteria for scores of 0, 1, and 2. IDENTIFY the key elements that distinguish a low score (0) from a high score (2).
   
2. Analyse the conversation in relation to specificity: ASSESS the extent to which the goal is specific rather than general. Are context and details of the goal explicitly described, and all terms explained? Is the goal concrete and attainable and not something abstract?

3. Assign a score:
Assign a score of 0, 1, or 2 based on the rubric. Use the provided scored examples as a reference to ensure consistency with previous assessments.

4. Provide a detailed rationale for the score:
EXPLAIN why you assigned the score by directly referencing aspects of the goal that meet or fall short of the rubric criteria.

5. Check for consistency:
Double-check that the score aligns with both the rubric criteria and the rationale provided. 
Maintain objectivity by strictly adhering to the rubric without introducing personal biases.

\#\#EDGE CASE HANDLING\#\#

If a goal is ambiguous or unclear, SCORE it on the lower end. If a goal appears to partially meet the criteria for two different scores, SELECT the score that best reflects the majority of the goals characteristics for that category.

\#\#WHAT NOT TO DO\#\#

Never apply personal opinion or assumptions outside the rubric criteria.
never give a score without a detailed explanation, even if the scoring seems obvious.
Never modify or assume student intent; score the goal exactly as written.
Never ignore the rubric or provided examples when scoring.
\end{quote}
\end{headerbox}

\begin{headerbox}{Running Example: Incorporating Few-shot Examples in System Prompt}
Use the following template to include few-shot examples.

\begin{quote}
\small
\#\#EXAMPLE SCORING\#\#

Example 1:
[insert example conversation]

Example 1 Scoring: [insert score] ([insert reasoning])

Example 2: 
[insert example conversation]

Example 2 Scoring: [insert score] ([insert reasoning])
\end{quote}
\end{headerbox}

\subsection{Prompt Templates and Automatic Generators}
Many LLM-powered chatbots can function as effective prompt-template generators. At a minimum, they provide useful starting points and inspiration. However, all generated prompts should be carefully reviewed and adjusted to the study's specific goals, data, and context.

\subsection{Summary}
Prompt engineering involves refining instructions to elicit higher-quality outputs from LLMs. A few straightforward techniques can substantially improve annotation quality: assigning the model a role; using unambiguous language and structure; providing labeled examples; and prompting step-by-step reasoning. LLM-powered chatbots can provide useful prompt templates, but these templates should always be adapted to the specific research context.

\section{Set Up the Computational Environment}
\label{sec:env}
\subsection{A Quick Start with Cloud Notebooks}
There are many valid setup options for working with LLMs, including local installations, cloud-based notebooks, virtual machines, and HPC clusters. For readers who are not experienced with R, Python, or programming in general, we recommend starting with \emph{cloud-based notebooks} such as Google~Colab (for Python and R) and RStudio~Cloud (for R). These platforms remove the need to manage local installations or hardware and are free at the basic tier. Our accompanying Python and R notebooks are designed to run on these services. Throughout this paper, the step-by-step demonstrations assume a Google~Colab session; however, the same commands can be executed locally with only minor adjustments. Once you are comfortable with this cloud-based notebook approach, we recommend moving toward a local, more repeatable and reproducible setup, especially for larger-scale text annotation projects (see \autoref{sec:reproducibility}).

\subsection{LLM APIs vs.\ Local LLMs}
Many people interact with LLMs daily through ``chat-based interfaces'' (e.g., ChatGPT). These interfaces are convenient for exploratory use because they allow users to ask questions in natural language and receive immediate responses. However, they are generally not well suited to research workflows. Chat interfaces make it difficult to systematically process large volumes of text, ensure reproducibility, control model parameters, or integrate models into larger automated pipelines. For research purposes, it is therefore usually preferable to interact with LLMs programmatically, typically in one of the following two ways:

\begin{itemize}
\item \textbf{Cloud APIs}, which provide access to hosted models via a remote service. These typically require only an internet connection and an API key, enabling easy access to both proprietary and open-source models depending on the provider.
\item \textbf{Local open-source models}, such as Llama and Qwen, which can be deployed directly on local computers, virtual machines, or high-performance computing (HPC) clusters.
\end{itemize}

For many researchers, cloud-based LLM APIs provide the most accessible entry point as they require minimal code to interact with and remove the need for dedicated (GPU) hardware. 
However, it is important to point out that this approach involves sending research data to the API provider, which may raise privacy and data protection concerns under legislation such as the GDPR. While some API providers claim not to store or process submitted data, annotating personal data using cloud APIs still requires explicit consent from study participants and must be covered by an approved data management protocol. In such cases, running local open-weight LLMs offers a feasible, GDPR-compliant, privacy-friendly alternative (see \autoref{sec:reproducibility}). This is especially true for scientific workflows where reproducibility, transparency, and long-term accessibility are central concerns. Compared to proprietary ones, open-weight LLMs can be downloaded, versioned, archived, and deployed locally or on institutional infrastructure, enabling annotation workflows that remain reproducible even if commercial APIs evolve or disappear.

In this paper, we use \texttt{Hugging Face Inference} as the default API provider for Python and \texttt{Groq} for R. Both host a wide selection of open-source LLMs. We further rely on the \texttt{langchain} and \texttt{ellmer} libraries to facilitate interaction with LLM APIs in Python and R, respectively. These libraries help keep the annotation pipeline provider-independent, such that the same workflow applies to other (proprietary) API providers and local setups.

\subsection{Obtain and Secure API Keys}
All cloud LLM providers require an API key linked to a user account. After creating an account on the provider's website, you can generate an API key in your user dashboard. The API key is a unique alphanumeric string that identifies you as the authorized user, for example, \texttt{c784df9e-bcf1-4145-83b6-a3c9ca281ee3}. This key grants access to the model and is used to monitor usage and manage billing. Note that the term ``API key'' is not adopted universally; some API providers use alternative terms like ``API/access/bear/authentication token'', ``personal access token (PAT)'', ``secrete key/token'' or ``API credential''. 

A useful practice is to generate separate API keys for different projects so you can better track the activities and costs associated with each project. To prevent accidental exposure of keys in shared repositories, never write them in plain text inside your scripts. Instead, store keys locally (and separately) in a Git-ignored file or as environment variables.

In the accompanying notebooks, the following code prompts a user to paste their key interactively; the key is retained only for that session. Both the Python and R code store the entered API key as an environment variable during a session to enable direct access by the LLM APIs.

\vspace{1em}
\noindent\textbf{Python Code} (Appendix~\ref{r_api_key} for R code):
\begin{minted}{Python}
# Prompt user for the API key securely (input is hidden) and store it for this session.
os.environ["HUGGINGFACEHUB_API_TOKEN"] = getpass.getpass("Enter API key: ")
\end{minted}

\subsection{Minimal Dependencies}
Install additional packages that the accompaning notebook require. For Python, that means a lightweight data stack and an API client; for R, an API client and tidy data tooling. 

\vspace{1em}
\noindent\textbf{Python Code} (Appendix~\ref{r_dependency} for R code):
\begin{minted}{python}
# Install additional packages not included in Colab by default
%pip install -qU langchain langchain-huggingface krippendorff
\end{minted}

\begin{minted}{python}
# Load required packages
from datetime import datetime  # Generates timestamps
import getpass  # For secure password or API key input (hides typed text)
import os  # For interacting with the operating system (e.g., paths, env vars)
import pandas as pd  # Data manipulation and analysis (tables, CSVs, etc.)
import numpy as np  # Numerical computing and array operations
from tqdm import tqdm  # Progress bars for loops and iterations
from langchain.chat_models import init_chat_model  # Initializes chat-based LLMs
from langchain_core.prompts import ChatPromptTemplate  # Template for building 
# structured LLM prompts
from langchain_huggingface import ChatHuggingFace, HuggingFaceEndpoint  # Imports hosted 
# Hugging Face endpoint wrappers.
from pydantic import BaseModel, Field  # Data validation and schema definition
import krippendorff  # Computes Krippendorff's alpha
\end{minted}

\subsection{Model Choice}
\paragraph{Keep it simple at the start.}
Different models involve trade-offs among \textit{accuracy}, \textit{cost}, and \textit{speed}, and the many choices can easily overwhelm users. To start, we recommend choosing a reasonably priced model recommended by the provider on its website.
In this paper, we use \texttt{Meta-Llama-3-8B-Instruct} for Hugging Face Inference and \texttt{llama-4-scout-17b-16e-instruct} for Groq. Both open-source models enable fast, affordable, and focused task completion.

\paragraph{Later on, keep costs in mind.}
Annotation projects involving tens of thousands of texts can incur substantial costs due to the large number of tokens consumed. To mitigate this, researchers should estimate total token usage by running a small pilot on a subset of the data. As a rough rule of thumb, one short paragraph ($\approx 100$ words) corresponds to approximately 120–150 tokens. Note that input tokens (i.e., the texts to be annotated) are often considerably cheaper than output tokens (i.e., the generated annotations). Overall, token costs have decreased over time as models become more efficient and competition among providers increases, although costs can still accumulate quickly at scale.

\subsection{Hyperparameters and Minimal Stability Settings}
LLM outputs can vary slightly between runs due to their stochastic nature. You can control this variance through the following commonly available hyperparameters:

\begin{itemize}
  \item \texttt{temperature}: sets output randomness (0 = deterministic, 1 = creative);
  \begin{itemize}
      \item Low values make the model more deterministic and repetitive. For example, a value of 0.3 was used for multiple-choice questions in GPT-4's exam benchmark~\citep{openai2024gpt4}. 
      \item High values increase diversity and creativity but may reduce coherence. For instance, 0.6 was selected for free-response questions in GPT-4's exam benchmark~\citep{openai2024gpt4}.
  \end{itemize}
  \item \texttt{top\_k} and \texttt{top\_p}: control sampling diversity.
  \begin{itemize}
      \item \texttt{top\_k}: restricts sampling to the $k$ most likely next tokens. A lower value (e.g., 10) makes output more deterministic. A higher value (e.g., 50 or 100) allows more diversity.
      \item \texttt{top\_p}: instead of selecting from the $k$ most probable tokens, this selects from the smallest set of tokens whose probabilities sum to $p$. Lower values (e.g., 0.3) make responses more deterministic. Higher values (e.g., 0.9) increase diversity.
  \end{itemize}
  \item \texttt{max\_tokens}: limits output length, which is especially useful for cost control;
  \item \texttt{seed}: fixes random initialization for reproducibility.
\end{itemize}

For example, to encourage consistent outputs from LLMs, you can set \texttt{temperature = 0} and fix a \texttt{seed}. Even so, a small degree of nondeterminism may remain due to factors such as server scheduling. Note also that not all hyperparameters are available for every LLM or API.
Always consult the official documentation to verify which hyperparameters are supported before use.

\subsection{Summary}
Begin in a cloud notebook to avoid setup friction, and use an API for a smooth first pass. Always keep your API keys secure. Set \texttt{temperature = 0} and fix a \texttt{seed} when reproducibility is desired, but postpone more detailed hyperparameter decisions until later.

\section{Run Your First Prompt Requests}
\label{sec:initial}

\subsection{Data Preparation}
Decide how you will store and access the text data to be annotated.
For most researchers, a simple \texttt{.csv} file is sufficient: it can be inspected with Excel, Numbers, or any text editor, and can be read easily by both Python and~R.
More efficient formats such as Parquet or Arrow can reduce file size and speed up I/O, but they are not necessary for small- or medium-sized projects.
Each row in the data should correspond to one text unit (e.g., a sentence, paragraph, or post), along with a unique identifier and any contextual metadata (e.g., participant ID, date, or source).
In this paper, we use a toy dataset from our running example of annotating goal specificity in student--chatbot conversations. The dataset contains the following columns:

\begin{headerbox}{Running Example: Dataset Description}

\begin{itemize}
  \item \texttt{id}: the row identifier (corresponding to one student and their entire conversation);
  \item \texttt{conversation}: the conversation text to be annotated;
  \item \texttt{best\_llm\_specificity\_score}: the final goal specificity score produced with a carefully engineered prompt by the best performing LLM, taking values 0, 1, or~2;
  \item \texttt{expert\_specificity\_score}: the gold-standard goal specificity score assigned by human expert annotators (also 0, 1, or~2);
  \item \texttt{grade}: the student's course grade (1--10 scale).
\end{itemize}

\end{headerbox}

To load the dataset:

\vspace{1em}
\noindent\textbf{Python Code} (Appendix~\ref{r_load_data} for R code):

\begin{minted}{python}
# Build the download URL for the example CSV dataset.
data_url = "https://sodascience.github.io/workshop_llm_data_collection/data/" \
    "srl_data_example.csv"

# Load the CSV from the URL into a pandas DataFrame.
df = pd.read_csv(data_url)

# Preview the first 15 rows
df.head(15)
\end{minted}

\subsection{A Single Prompt Demo}
\label{subsec:simple_system_prompt}
Below, we demonstrate how to send a single API prompt and obtain a response. First, we create a simple system prompt describing the task, context, and output.

\vspace{1em}
\noindent\textbf{Python Code} (Appendix~\ref{r_single_prompt} for R code):
\begin{minted}{python}
# Build the system prompt
system_prompt = """
You are an expert in educational assessment and goal evaluation, with
specialized expertise in applying deductive coding schemes to score the quality
and content of student goals.

##TASK##
A university student was given a series of prompts, guiding them through the
process of setting and elaborating on an academic goal for the coming week. You
will be provided with the entire conversation including the prompts, and the
student answers. Your objective is to assess the specificity of of the student's
goal on a scale of 0 to 2 based on the entire conversation.
"""
\end{minted}

Second, we use LangChain's prompt-template module to create a request containing both a system prompt and a user prompt.

\begin{minted}{Python}
# Define a two-part prompt template: fixed system instructions + user input.
prompt_template = ChatPromptTemplate([
    ("system", system_prompt),
    ("user", "{conversation}")
])

# Populate the {conversation} placeholder with the first conversation example.
prompt_request = prompt_template.invoke({"conversation": df.iloc[0,1]})

# Inspect the rendered messages before sending the request to the model.
prompt_request.to_messages()
\end{minted}

Finally, initialize your model with the desired hyperparameter settings, run the prompt, and inspect the response.

\begin{minted}{python}
# Initialize the chat model with deterministic settings for reproducibility.
hf_endpoint = HuggingFaceEndpoint(
    repo_id="meta-llama/Meta-Llama-3-8B-Instruct",
    temperature=0,
    max_new_tokens=1000,
    seed=123,
)
model = ChatHuggingFace(llm=hf_endpoint)

# Send the prepared prompt to the model.
response = model.invoke(prompt_request)

# Print the model's text output.
print(response.content)
\end{minted}

Voil\`a! You have completed your first successful interaction with an LLM API.

\begin{headerbox}{Running Example: First Prompt Results}
Below is an example response generated by the single-prompt demo code:

\begin{quote}
\small
To assess the specificity of the student's goal, we can evaluate the details provided throughout the conversation.

1. **Initial Goal**: The student starts with a broad goal of "catching up on my geography reading." This is quite vague and does not specify what exactly needs to be done.

2. **Added Details**: The student then elaborates by stating they need to either read the book from last week and this week or read their friend's notes. This adds some specificity, as it identifies the materials to be covered (the book and friend's notes) but still lacks a clear plan on how much reading will be done.

3. **Measurement of Progress**: The student mentions measuring progress by "the number of pages I write per day." This is a measurable aspect, but it does not specify how many pages are the target or how this relates to the overall goal of catching up.

4. **Importance of the Goal**: The student explains the importance of the goal in relation to not falling behind and being prepared for the exam. This adds context but does not enhance the specificity of the goal itself.

5. **Step-by-Step Plan**: The student provides a three-step plan:
   - Evaluate how much there is to do
   - Get help from friends
   - Take notes day by day
   This plan outlines steps but remains somewhat general. It does not specify timelines or the amount of reading or note-taking expected each day.

Based on these observations, the goal has moved from a vague statement to a more defined plan, but it still lacks concrete details regarding the amount of reading, specific timelines, and measurable outcomes. 

**Score**: 1 (The goal is somewhat specific but still lacks clarity and measurable targets.)
\end{quote}
\end{headerbox}

\subsection{Multiple Prompts Demo}
\label{subsec:multiple_prompts}
Next, we show how to process multiple prompts using a \texttt{for} loop. To do this, we first define a list of request IDs and a corresponding list of conversations (used to form the user prompts). For demonstration purposes, we use the first 10 conversations in the dataset.

\vspace{1em}
\noindent\textbf{Python Code} (Appendix~\ref{r_multi_prompts} for R code):

\begin{minted}{Python}
# Select the first 10 request IDs and conversations for a small demo batch.
ids = df.id[:10].tolist()
conversations = df.conversation[:10].tolist()

# Store outputs in a dictionary keyed by request ID.
responses = {}
for id, conversation in tqdm(zip(ids, conversations),
                             total=len(ids),
                             desc="Processing Requests"):
    # Insert each conversation into the prompt template.
    prompt_request = prompt_template.invoke({"conversation": conversation})
    # Send the prompt to the model and collect the response.
    response = model.invoke(prompt_request)
    # Save only the text content for later analysis.
    responses[id] = response.content

# Print one example response for quick inspection.
print(responses['chat_3'])
\end{minted}

\section{Structure LLM Annotations}
\label{sec:struct}

LLM responses are typically returned as free-form text, which complicates their use as variables in downstream analyses. In our running example, where we ask the LLM to assign a score from 0 to 2, the model may produce outputs such as “two,” “this conversation scores a 2,” and other variants. Extracting a numeric score from such responses requires additional parsing, which increases the risk of errors and adds substantial processing time. To address this issue, we can leverage a technique called \emph{structured output}, which enforces user-specified formats for model responses. In our running example, this means requiring each LLM response to contain exactly one \emph{integer} goal specificity value between 0 and 2 and, optionally, a string field for reasoning.
Furthermore, combining the workflow in \hyperref[subsec:multiple_prompts]{\ref*{subsec:multiple_prompts}: \nameref*{subsec:multiple_prompts}} with structured output can save substantial time in research projects.

To force the LLM to produce outputs in user-specified formats, you can use the \texttt{BaseModel} and \texttt{Field} classes from the \texttt{pydantic} package.
Below, we define our desired output format as:

\begin{itemize}
    \item \texttt{specificity\_score}: an integer (0, 1 or 2) reflecting the goal specificity of a conversation.
    \item \texttt{reasoning}: a string that provides the model's reasoning.
\end{itemize}

\vspace{1em}
\noindent\textbf{Python Code} (Appendix~\ref{r_structured_output} for R code):

\begin{minted}{Python}
# Define the schema that the model must follow in every response.
class SpecificityFormat(BaseModel):
    """Structured output format for goal specificity annotation."""
    specificity_score: int = Field(
        description="Score for goal specificity. Only return an integer from 0 to 2."
    )
    reasoning: str = Field(description="The reasoning to justify the score")


# Attach the schema to the base model to enforce structured outputs.
model_structured = model.with_structured_output(SpecificityFormat)

# Create dictionaries to store parsed fields per request ID.
structured_scores = {}
structured_reasonings = {}
for id, conversation in tqdm(zip(ids, conversations), total=len(ids), desc="Processing"):
    # Render the prompt with the current conversation.
    prompt_request = prompt_template.invoke({"conversation": conversation})
    # Get a schema-compliant response (score + reasoning).
    structured_response = model_structured.invoke(prompt_request)
    # Save each field separately for downstream analysis.
    structured_scores[id] = dict(structured_response)["specificity_score"]
    structured_reasonings[id] = dict(structured_response)["reasoning"]

\end{minted}

This code works with many API providers. However, using the Hugging Face Inference API requires some additional adjustments because structured output validation must be specified explicitly. The accompanying notebooks provide the extra code needed for this setup.

\begin{headerbox}{Running Example: Structured Output Results}
Below is an example of a structured response in Python dictionary format:

\begin{quote}
\small
\textbf{specificity\_score}: 1,

\textbf{reasoning}: ``The student's goal of catching up on geography reading is somewhat specific, as they mention reading the book from the previous week and this week or using friends' notes. However, the goal lacks precise details such as the exact number of pages to read or a specific timeline for completing the readings. The measurement of progress is vague ('number of pages I write per day') and does not provide a clear target. The step-by-step plan includes some actionable items but could benefit from more specificity regarding deadlines and quantifiable outcomes.''
\end{quote}
\end{headerbox}

\section{Evaluate Annotation Quality}
\label{sec:evaluation}
A crucial step in evaluating the quality of LLM-based annotations is quantifying their level of agreement with a reference (or ``gold-standard'') dataset. The choice of agreement metric depends on the nature of the annotation outcome (i.e., categorical, ordered, or continuous).

\subsection{Apply Common Agreement Metrics}
\paragraph{Categorical outcomes.}
For nominal labels (e.g., news categories such as \emph{politics}, \emph{science}, and \emph{art}), the most commonly used metrics include simple accuracy, Cohen's~$\kappa$~\citep{cohen1960}, and Krippendorff's~$\alpha$~\citep{krippendorff2018}.
Accuracy is intuitive and easy to interpret, but it ignores chance agreement and is sensitive to class imbalance (i.e., when categories are rare). Therefore, accuracy is typically used only with balanced datasets. Cohen's~$\kappa$ adjusts for chance agreement between two raters, providing a more conservative estimate, while Krippendorff's~$\alpha$ generalizes this idea to multiple annotators and handles missing data.

\paragraph{Ordered outcomes.}
When labels are ordinal (e.g., Likert-type ratings), weighted versions of $\kappa$ or $\alpha$ are preferred because they account for the magnitude of disagreement between categories. Linear or quadratic weights penalize larger discrepancies more heavily, thereby reflecting the underlying ordering of the labels~\citep{gwet2014handbook}.

\paragraph{Continuous outcomes.}
For (quasi-)continuous\footnote{Quasi-continuous variables are discrete variables (often ordinal or interval-like) with many possible values that are treated as continuous for practical or statistical reasons.} variables (e.g., predicted probabilities, test scores, and frequencies), agreement is typically quantified using correlation-based or error-based metrics.
The Pearson correlation coefficient ($r$) assesses the strength of linear association, while the intraclass correlation coefficient (ICC)~\citep{shrout1979intraclass} measures both correlation and absolute agreement.
Alternatively, mean absolute error (MAE) or root mean squared error (RMSE) can be used to express the average discrepancy in scale units.

\paragraph{Note.}
In practice, no single metric fully captures annotation quality. A comprehensive evaluation often combines multiple indicators to balance interpretability and robustness, e.g., reporting both $\kappa$ and class-specific precision/recall for categorical data, or both ICC and MAE for continuous data.
Additionally, these statistics should be complemented by a qualitative inspection of disagreements, which we discuss next in \autoref{subsec:problematic_items}.

Below is a code implementation of Krippendorff's~$\alpha$ to evaluate agreement between LLM annotations and gold-standard labels in our running example.

\vspace{1em}
\noindent\textbf{Python Code} (Appendix~\ref{r_annotation_quality} for R code):
\begin{minted}{Python}
def compute_krippendorff_alpha(x: list[int], y: list[int]):
  # Format data into a reliability matrix (rows=raters, cols=items)
  data_krippendorff = np.array([x, y])
  # Compute Krippendorff's Alpha (interval metric)
  kripp_alpha = krippendorff.alpha(reliability_data=data_krippendorff, 
  level_of_measurement='interval')
  return kripp_alpha

# Check the agreement between the specificity scores between LLM and human experts
expert_specificity_scores = df.expert_specificity_score[:10].tolist()
structured_llm_specificity_scores = list(structured_scores.values())
print("Krippendorff's Alpha:", 
compute_krippendorff_alpha(structured_llm_specificity_scores, expert_specificity_scores))
\end{minted}

\begin{headerbox}{Running Example: Annotation Quality}
Using Krippendorff's~$\alpha$, we evaluate the performance of the system prompt defined in \autoref{subsec:simple_system_prompt} on a sample of 10 conversations. The resulting score is 0.222, indicating low agreement, poor prompt or gold-label quality, and a need for further inspection.
\end{headerbox}

\subsection{Identify Problematic Items and Examine Issues}
\label{subsec:problematic_items}
Regardless of the overall agreement score, it is essential to examine and, where possible, correct potentially problematic items. Quantitative agreement metrics can mask specific sources of error, so visual inspection of cases with disagreements between the LLM and the gold labels is highly recommended.
Such inspection may reveal that (a) the gold annotation itself is incorrect due to human oversight; (b) the LLM output violates task specifications, for example by producing a value outside the permitted range; or (c) the LLM fails to account for certain nuances, exceptions, or contextual cues described in the prompt. Identifying these patterns not only improves label quality but also clarifies whether disagreement stems from annotation ambiguity, model behavior, or flaws in prompt design.

\subsection{Fix Data Issues and Improve Prompts Iteratively}
Once potential data issues have been identified, the next step is to iteratively refine both the data and the prompting procedure.
This process typically involves three complementary actions: (1) fixing the data itself, (2) improving the prompt (or sometimes the research task specification itself), and (3) re-running the annotation workflow to verify improvements.

\paragraph{1.~Correct or flag problematic items.}
If disagreements stem from clear human or LLM errors (e.g., invalid label values, missing entries, or mismatched examples), these cases should be corrected or flagged for exclusion.
When corrections are uncertain (e.g., due to human disagreements), it is good practice to keep both the original and revised versions in a transparent audit trail, noting the rationale for each change.
In projects with multiple annotators, such cases can also be reviewed by an adjudicator or majority vote to ensure consistency.

\paragraph{2.~Refine prompt instructions.}
When disagreements reflect systematic LLM misinterpretations, the prompt itself likely needs revision.
Common improvements include clarifying category definitions, adding boundary examples, explicitly stating how to handle ambiguous or multifaceted cases, and enforcing stricter output formatting.
For instance, if the LLM tends to overestimate goal specificity scores, the prompt can be refined with counterexamples (e.g., ``Responses that only restate the goal without a concrete plan should receive a 0'').
Iterative prompt adjustments are particularly effective when combined with structured outputs (e.g., JSON schema) that enforce the required response format.

\paragraph{3.~Re-run and re-evaluate.}
After each refinement round, re-run the annotation and re-calculate agreement metrics to assess improvement.
A well-documented iteration log, including date, prompt version, and agreement statistics, allows researchers to track progress and determine whether changes in agreement reflect genuine prompt improvement or random variation.
When multiple prompt versions are tested during the iterative development process, their performance can be tracked in a comparison table or plot.
More systematic prompt comparison and selection are discussed further in \autoref{sec:best_prompt}.

\subsection{Summary}
Evaluating LLM-based annotations is not a one-off step but a cyclic process of measurement, diagnosis, and improvement.
Quantitative agreement metrics such as $\kappa$, $\alpha$, and ICC provide essential summaries of annotation quality, yet they must be interpreted alongside qualitative inspection of disagreements and iterative prompt adjustments.
Poor agreement does not necessarily imply model failure; it often signals ambiguous instructions, unclear category boundaries, or inconsistencies in the gold data itself.
An effective evaluation workflow therefore combines agreement assessment, error diagnosis, and iterative refinement.
Through this systematic and transparent approach, researchers can ensure that LLM annotations not only replicate human judgments but also reveal where human definitions and instructions may need clarification, turning evaluation into a process of both quality control and conceptual refinement.

\begin{headerbox}{Running Example: Iterative Refinement}

For the 10 evaluated conversations, the (LLM-generated, gold) label pairs are:
\begin{quote}
\small
$(1,1)$, $(2,2)$, $\textbf{(2,1)}$, $(1,0)$, $(2,1)$, $(2,1)$, $(2,1)$, $(1,0)$, $(2,1)$, $(2,2)$.
\end{quote}

The LLM-provided reasoning for the first disagreement case (the third item) is: 

\begin{quote}
\small
The student's goal is specific, measurable, and actionable. They clearly state the intention to start studying right after class, specify the action of reading a quarter of the readings, and outline a step-by-step plan to achieve this. The goal is also tied to their personal motivation to combat procrastination, which adds depth to its importance. Overall, the goal meets the criteria for high specificity.
\end{quote}

Among others, two strategies can be used to examine disagreements between the LLM-generated and gold-standard labels. \emph{First}, inspecting the overall pattern of disagreement reveals that the LLM tends to systematically overestimate goal specificity scores across conversations. This pattern suggests potential issues with the scoring instructions in the prompt. Incorporating few-shot examples may help calibrate the model and reduce systematic over- or underestimation.
\emph{Second}, examining individual cases can provide more fine-grained diagnostic insights. For instance, in the third conversation, the LLM justifies its label by stating that the goal is ``specific, measurable, and actionable.'' This indicates that the model is not only assessing goal specificity, but also incorporating measurability and actionability: two conceptually distinct dimensions of self-regulated learning. This points to a likely underdefinition of goal specificity in the system prompt.
By iteratively diagnosing such discrepancies and refining the system prompt, we eventually achieve a Krippendorff's~$\alpha$ of 0.894. The final prompt is provided in \autoref{final_prompt}. \emph{Finally}, note that this example does not assess the quality of the gold-standard labels themselves.
\end{headerbox}

In practice, the procedures described in this section are best applied to a \emph{development subset} of the gold-labeled data.
This allows researchers to inspect errors freely, revise prompts, and iterate on the annotation workflow without affecting the validity of the final performance estimates.
If the same labeled dataset is repeatedly used both to refine prompts and to report agreement metrics, the resulting performance estimates can become overly optimistic.
The next section (\autoref{sec:best_prompt}) explains why this issue arises and introduces a framework that separates prompt exploration, prompt selection, and final evaluation, allowing researchers to iteratively improve prompts while maintaining valid and unbiased performance estimates.

\section{Search for the Best Prompt}
\label{sec:best_prompt}
\autoref{sec:evaluation} introduced an iterative workflow for diagnosing errors and improving prompts by inspecting disagreements between LLM annotations and gold-standard labels.
This iterative process is extremely crucial: examining problematic cases, refining instructions, and re-running annotation often yields substantial improvements in annotation quality.

However, there is one major caveat if this procedure is repeatedly applied to the \emph{entire} gold-labeled dataset.
By repeatedly adjusting prompts based on the same labeled examples used to measure agreement, the prompt design becomes increasingly tailored to the specific characteristics of that dataset.
As a result, measured performance may appear better than the prompt would actually perform on new data.

This phenomenon is commonly known as \emph{overfitting}.
In simple terms, overfitting occurs when a model or procedure adapts too closely to the quirks of the data used during development.
Consequently, its apparent performance becomes overly optimistic when evaluated on the same dataset that guided its design.
In the context of LLM-based annotation, overfitting arises because researchers select among multiple prompt designs using performance estimates computed on the same labeled dataset.
Because this dataset represents only a limited sample of texts from a broader population, the measured agreement scores for each prompt may vary slightly depending on which specific items appear in the dataset.
When researchers choose the prompt with the highest observed score on this dataset, they may inadvertently select one that benefited from favorable sampling variation rather than one that truly performs better.

To address this issue, we suggest a simple three-stage framework for prompt development and evaluation: \textbf{Explore} $\rightarrow$ \textbf{Select} $\rightarrow$ \textbf{Evaluate}.
The key principle underlying this framework is the separation between \emph{prompt development} and \emph{performance evaluation}.
Any labeled data used to improve or select prompts should not be used to report final annotation quality.

\subsection{Why Prompt Selection Can Produce Biased Performance Estimates}
Suppose a researcher evaluates several candidate prompts on a labeled dataset and chooses the prompt with the highest agreement score.
Even if several prompts have identical true performance, their measured scores on a finite labeled dataset will vary slightly because the dataset represents only a sample of texts from a broader population.
Selecting the prompt with the highest observed score therefore tends to favor prompts that benefited from favorable sampling variation rather than genuinely superior instructions.

This effect is sometimes referred to as the \emph{winner's curse} in statistics and the model--selection literature \citep{cawley2010overfitting}.
The key point is that the bias arises from \emph{selection based on the same data used for evaluation}.
Importantly, this phenomenon does not depend on model training.
It occurs whenever researchers compare multiple alternatives using the same labeled dataset, including prompt engineering, feature selection, and questionnaire design.

Readers familiar with cross-validation in classical machine learning may wonder how these ideas apply in the context of LLM-based annotation, where the model itself is not trained on the data.
In traditional supervised learning, cross-validation serves two purposes: selecting hyperparameters (such as regularization strength or model architecture) and estimating model performance.
Because model parameters are re-estimated on different subsets of the data, the resulting fitted models differ across folds.
Nested cross-validation is therefore often used to separate these two tasks: inner folds are used to select hyperparameters, while outer folds provide an unbiased estimate of the performance of the resulting trained model \citep{cawley2010overfitting,varma2006bias}.

In prompt-based annotation, however, the situation is slightly different.
The underlying LLM parameters remain fixed; prompt design only changes the instructions given to the model.
As a result, evaluating the same prompt on different folds would typically produce the same predictions for each item (assuming deterministic decoding), which means that nested cross-validation does not provide the same benefits it does for models whose parameters are re-trained on each split.
Nevertheless, prompt selection still resembles hyperparameter tuning in classical machine learning: researchers compare several prompt candidates using labeled data and choose the one with the best observed performance.
This selection process can still lead to overly optimistic performance estimates if the same labeled data are used both to choose the prompt and to report its performance.
\emph{The purpose of separating development and evaluation datasets in prompt engineering is therefore not to prevent the model from learning the training data, but to ensure that the final performance estimate is not biased by the prompt-selection process.}

\subsection{The Explore->Select->Evaluate Framework}
To reduce bias in performance estimation while preserving the benefits of iterative prompt development, we recommend dividing the labelled data into three subsets, each aligned with a stage of the workflow.

\begin{enumerate}
\item \textbf{Explore (prompt development).}  
A small portion (10--20\%) of the labelled data is used for exploratory prompt design.
Researchers may freely inspect annotation errors, modify instructions, add examples, and iteratively refine prompts, following the procedures described in \autoref{sec:evaluation}.
Overfitting at this stage is expected and acceptable because the objective is to identify obvious wins and failures in prompt engineering strategies, on this basis, generate a set of promising prompt candidates.

\item \textbf{Select (prompt comparison).}  
A larger validation subset (40--60\%) is used to compare a fixed set of candidate prompts produced at the end of the exploration stage.  
All candidate prompts are evaluated on exactly the same labelled examples, and the prompt with the best agreement metric is selected.  
After inspecting performance results on this subset, no further prompt modifications should be made.

\item \textbf{Evaluate (final performance estimation).}  
The final prompt is evaluated on a held-out test subset (20--30\%) that has not been used during prompt development or comparison.  
This ensures that reported agreement statistics provide an unbiased estimate of annotation quality on unseen data.
\end{enumerate}

\subsection{Relationship to the Iterative Workflow in \autoref{sec:evaluation}}
The iterative procedure described in \autoref{sec:evaluation} naturally fits within the \textbf{Explore} stage of the framework.  
Inspecting disagreements, revising prompts, and re-running annotations are essential for improving task instructions and identifying conceptual ambiguities.  
However, these iterative adjustments should be restricted to the exploration subset of the labelled data.
Once researchers move to the \textbf{Select} stage, the set of candidate prompts should be frozen before evaluation.  
Allowing further prompt modifications after observing validation results would reintroduce the same selection bias that the framework is designed to avoid.
Finally, the \textbf{Evaluate} stage provides the performance estimates to be reported in the study.  
At this point, both the prompt design and the LLM hyperparameters should remain fixed.

\subsection{Handling Stochastic LLM Outputs}
When stochastic decoding is used (e.g., temperature sampling), the same prompt may produce different annotations across runs.  
To obtain stable estimates, researchers can repeat the annotation process multiple times and average the resulting agreement metrics.  
This approach estimates the expected performance of the prompt while keeping the evaluation dataset fixed.

\subsection{Summary}

Prompt engineering inevitably involves iterative experimentation and data-driven decision-making.  
Consequently, some degree of overfitting during development is unavoidable.  
The goal of a rigorous workflow is therefore not to eliminate overfitting entirely, but to prevent it from biasing the final performance estimates.
The Explore $\rightarrow$ Select $\rightarrow$ Evaluate framework provides a simple structure for achieving this goal.  
By clearly separating exploratory prompt development, prompt selection, and final evaluation, researchers can iteratively improve annotation quality while maintaining valid and transparent performance estimates.

\section{Consider Annotation Error in Downstream Analyses}
\label{sec:downstream}
While agreement metrics provide an indication of annotation quality, researchers should also consider how both random and systematic annotation error can affect downstream analyses. Ignoring such error may lead to biased parameter estimates, attenuated relationships, and incorrect inferences. This can occur even when accuracy and inter-rater agreement are high~\citep{egami2023dsl}.

\subsection{Systematic Error vs. Random Error in Annotations}
Errors in annotations (or any measurements) can broadly be classified as either \emph{random} or \emph{systematic}.  
Random error refers to unsystematic variation in annotation outcomes that arises from factors such as ambiguity in the text, inconsistent judgments by human annotators, or stochastic variation in LLM outputs when sampling-based decoding is used. Because these errors occur unpredictably across items, they introduce variability in the produced labels but do not consistently push annotations in a particular direction.

Systematic error, by contrast, occurs when annotations deviate from the intended construct in a consistent and predictable way.  
For example, an LLM might systematically assign higher scores than do human annotators, consistently misclassify certain topics, or perform worse for specific linguistic styles or demographic groups due to biases in its training data. In such cases, the annotation process produces labels that are systematically distorted relative to the conceptual definition of the variable being measured.

\subsection{Effects on Downstream Analyses}
Distinguishing between these two sources of error is important because they affect downstream analyses in different ways. Random error primarily reduces the \emph{reliability} of measurements by introducing additional variability \citep{carmines1979reliability}. In statistical analyses, this typically attenuates estimated relationships and reduces statistical power, making it harder to detect real effects \citep{fuller1987measurement}. Systematic error, in contrast, threatens \emph{validity}: if annotations consistently misrepresent the underlying construct, statistical analyses based on these labels may produce biased or misleading conclusions even in large samples \citep{fuller1987measurement}.

Below, we provide intuitive explanations of how systematic and random error in both predictors and outcomes can affect mean estimation and regression results (e.g., coefficients, standard errors, and \textit{p}-values) based on a simple linear regression example.

\subsection{A Simple Linear Regression Example}

\begin{figure}[ht]
    \centering
    \includegraphics[width=\linewidth]{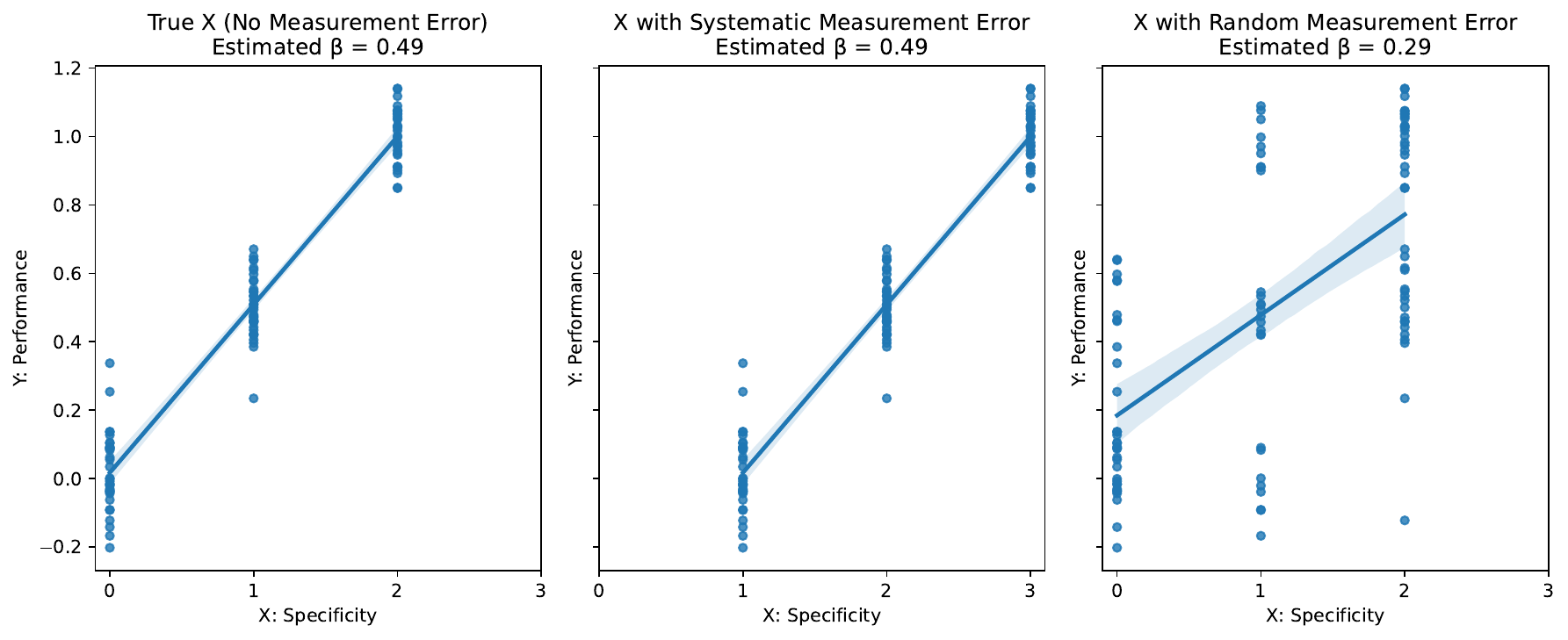}
    \caption{Illustration of how systematic error and random error in the predictor (X) of a simple linear regression analysis affect model estimates.}
    \label{fig:measurement_error}
\end{figure}

Figure~\ref{fig:measurement_error} illustrates how the two types of annotation error affect statistical estimates in a simple linear regression setting, where student performance scores (Y) are regressed on goal specificity scores (X). The data used to generate this figure were simulated and intentionally designed to demonstrate the qualitative effects of measurement error. Due to space constraints, the code used to generate these simulations and figures is not included in the paper, but it is available in the accompanying project notebooks.

In the left panel, the predictor $X$ is measured without error. The scatterplot therefore reflects the true relationship between $X$ and the outcome $Y$, and the estimated regression slope ($\hat{\beta} = 0.49$) closely recovers the true underlying association ($\beta = 0.50$).

The middle panel shows a case in which $X$ contains \emph{systematic measurement error}. Here, the observed values of $X$ are consistently shifted upward by one unit, but the ordering of observations remains unchanged. As a result, the estimated regression slope remains essentially the same as in the error-free case ($\hat{\beta} = 0.49$). In other words, when measurement error simply adds a constant or otherwise preserves the relative ranking of observations, the estimated relationship between $X$ and $Y$ may remain largely unaffected. However, as the figure shows, the previously positive intercept has become negative.

More importantly, even if the regression slope is unchanged, the interpretation of the measured values becomes problematic when the true labels are unobserved. In this example, the goal specificity scores were intended to take values in $\{0,1,2\}$, yet the observed measurements now include values up to $3$. Without access to the true labels, it becomes unclear how these values should be interpreted. Should all the scores be shifted down by 1 (treating $3$ as $2$, $2$ as $1$, and $1$ as $0$)? Or do the measurements correctly identify cases with scores $1$ and $2$, with the apparent shift indicating that there are simply no true $0$ observations in the sample? Such ambiguity reveals a deeper problem with the measurement instrument (LLM or human annotator): although the relative ordering of observations is preserved, the observed scores no longer correspond transparently to the intended scale of the underlying construct.

The right panel illustrates \emph{random measurement error} in $X$. In this case, the observed values of $X$ vary unpredictably around their true values. This additional noise weakens the observed association between $X$ and $Y$, causing the regression slope to shrink toward zero ($\hat{\beta} = 0.29$) with a larger uncertainty interval. This phenomenon is known as \emph{attenuation bias} \citep{fuller1987measurement}. In practical terms, random annotation error makes real relationships harder to detect because the predictor becomes a noisier proxy for the underlying construct.


\subsection{Software Packages for Correcting Annotation Error}
Several statistical methods and software packages have been developed to account for annotation/measurement error when using machine-generated labels (e.g., LLM annotations) in downstream statistical analyses.
These methods typically combine a relatively small set of gold-standard observations (denoted $Z$) with a larger set of machine-generated annotations (denoted $\hat{Z}$).
The gold-labelled subset provides information about prediction or annotation error, which can then be used to correct downstream statistical estimates.

Conceptually, existing approaches differ in \emph{where} the correction is applied in the analysis pipeline.
Some methods attempt to correct the generated labels $\hat{Z}$ themselves by modeling their relationship with the gold-standard labels $Z$.
Other approaches instead modify the \emph{loss function}, or the estimating equation used to estimate model parameters, so that label error is accounted for during model fitting.
A third possibility is to directly correct the \emph{final statistical estimates} after model estimation using closed-form adjustments.
We focus on the first two approaches, as they are more flexible and currently the most widely implemented in available software tools.

\paragraph{Correct $\hat{Z}$}
These methods model the relationship between machine-generated labels $\hat{Z}$ and gold-standard labels $Z$ using the labelled subset.
For example, \emph{Post-Prediction Inference} estimates a statistical model predicting $Z$ from $\hat{Z}$ and then uses this relationship to correct downstream estimators \citep{wang2020postpi}.
Similarly, \emph{Prediction-Powered Inference} models the prediction error $(\hat{Z} - Z)$ using auxiliary covariates $W$ observed in the labelled subset, which can then be used to construct unbiased estimators with improved efficiency \citep{angelopoulos2023prediction}.
\emph{Design-based Supervised Learning} takes a related approach but incorporates sampling-weight corrections when estimating $Z$ from $\hat{Z}$ and $W$~\citep{egami2023dsl}.
More recently, \emph{Assumption-Lean and Data-Adaptive Post-Prediction Inference} extends this class of approaches by flexibly estimating the relationship between $Z$ and $\hat{Z}$ with minimal modeling assumptions, allowing the correction to adapt to complex prediction-error structures \citep{miao2025pspa}.

\paragraph{Correct loss function}
A second class of approaches corrects the \emph{loss function} used to estimate model parameters.
Rather than explicitly correcting $\hat{Z}$, these methods modify the estimating equation so that the influence of prediction or annotation error is removed or adjusted during model estimation.
For instance, \emph{Task-Agnostic Machine-Learning-Assisted Inference} proposes a framework for constructing such corrections directly at the level of the estimating equation, allowing valid inference for a broad class of downstream estimators without requiring task-specific adjustments \citep{miao2024psps}.
\emph{Prediction-Powered Inference} can also be interpreted as adding a correction term derived from the observed error $(\hat{Z}-Z)$ in the labelled subset to the loss function used to estimate the target parameter.

\paragraph{Software Overview}
Several software packages implement these approaches.
For example, the \texttt{postpi} R package implements Post-Prediction Inference for estimators such as means, quantiles, and generalized linear models.
The \texttt{ppi\_py} Python package implements Prediction-Powered Inference and several extensions.
The \texttt{pspa} R package implements \emph{Assumption-Lean and Data-Adaptive Post-Prediction Inference} and can handle label error in both predictors and outcomes.
The \texttt{dsl} R package implements \emph{Design-based Supervised Learning} and can also correct both predictors and outcomes.
The \texttt{psps} package is supported in both Python and R and implements \emph{Task-Agnostic Machine-Learning-Assisted Inference}.
Finally, the \texttt{ipd} package provides a unified interface for several of these tools, including \texttt{postpi}, \texttt{ppi}, and \texttt{pspa} \citep{salerno2025ipd}.

Table~\ref{tab:annotation_error_tools} summarizes these software tools, the statistical estimators they support, and whether they correct label error in outcomes, predictors, or both.

\begin{table}[htbp]
\footnotesize
\centering
\caption{Software packages for correcting annotation error in downstream statistical analyses.}
\label{tab:annotation_error_tools}
\begin{tabularx}{\textwidth}{l l l l}
\toprule
\textbf{Name} & \textbf{Language} & \textbf{Supported Estimators} & \textbf{Corrects} \\
\midrule

\href{https://github.com/leekgroup/postpi}{postpi} 
& R 
& Means, quantiles, generalized linear models 
& Outcome \\

\href{https://github.com/aangelopoulos/ppi_py}{ppi\_py} 
& Python 
& Arbitrary estimators 
& Outcome \\

\href{https://github.com/qlu-lab/pspa}{pspa} 
& R 
& Means, quantiles, linear regression, logistic regression 
& Predictor \& outcome \\

\href{https://naokiegami.com/dsl/}{dsl} 
& R 
& Moment-based estimators 
& Predictor \& outcome \\

\href{https://github.com/qlu-lab/psps}{psps} 
& R, Python 
& M-estimators 
& Outcome \\

\href{https://github.com/ipd-tools/ipd}{ipd} 
& R 
& Means, quantiles, linear regression, logistic regression 
& Outcome \\

\bottomrule
\end{tabularx}
\end{table}

Our accompanying notebooks demonstrate two practical implementations:

\begin{itemize}
  \item \textbf{Python:} The \texttt{ppi\_py} package provides functions such as \texttt{ppi\_ols\_pointestimate()} and \texttt{ppi\_ols\_ci()} for correcting regression coefficients and confidence intervals under classical measurement error assumptions.
  \item \textbf{R:} The \texttt{dsl} package implements a declarative syntax for specifying measurement models and performing bias correction in regression and structural equation settings.
\end{itemize}

Together, these approaches enable researchers to quantify and, where possible, mitigate the influence of annotation error on downstream statistical analyses, thereby improving the validity and interpretability of LLM-assisted analyses.

\begin{headerbox}{Running Example: Regression Correction}
A detailed illustration of the influence of systematic and random error in the predictor variable on a simple linear regression outcome, along with demonstrations of both the \texttt{ppi\_py} and \texttt{dsl} packages, is available in our accompanying tutorial notebooks.
\end{headerbox}

\subsection{Summary}
When annotations are used as variables in downstream analyses, annotation error can substantially influence both descriptive statistics and inferential statistical analyses.
Random error primarily reduces reliability, inflates variance, and attenuates observed associations, while systematic error introduces directional bias that can distort estimated means and substantive interpretations.
To ensure valid conclusions, researchers should assess annotation quality, examine potential sources of systematic bias, and apply appropriate statistical correction tools.







\section{Take Your Annotations to the Next Level}
\label{sec:nextlevel}

Once the basic annotation workflow is in place, researchers can further improve the \emph{efficiency}, \emph{performance}, and \emph{reproducibility} of their LLM-assisted annotation pipelines. This section outlines several best practices and advanced strategies, as well as common pitfalls to avoid.

\subsection{Efficiency}

\paragraph{Caching.}
When running iterative or large-scale annotation tasks, repeated prompts can lead to unnecessary API calls and avoidable costs. To minimize redundancy, keep the initial tokens of the system prompt (i.e., the instructions and context that precede the variable content) identical across calls. This allows some API providers to cache and reuse shared prompt context, reducing both latency and cost.

\paragraph{Batch processing.}
For large datasets, asynchronous batch submission can be considerably cheaper than sending prompts sequentially. Some API providers support this approach, allowing researchers to upload a file containing all requests and retrieve the results within a fixed time window at a substantially reduced rate (e.g., a 50\% cost reduction and results within 24 hours with OpenAI's API).

\subsection{Performance}

\paragraph{Model hyperparameter tuning.}
The most influential hyperparameter for annotation quality is \texttt{temperature}. Lower temperatures (e.g.,~0 to~0.3) produce more deterministic outputs and are typically recommended for classification or coding tasks. Higher values increase variability and may encourage creativity, but at the cost of consistency.
Where possible, evaluate multiple temperature settings using a small validation subset to identify the best temperature value for your specific task.

\paragraph{Consistency prompting.}
LLMs can exhibit stochastic variation even under identical prompts. A simple yet powerful way to increase reliability is to generate multiple responses for each item and aggregate the results, for example, by taking the majority vote for categorical outputs or the mean for continuous scores. This ensemble-like strategy smooths out random fluctuations and provides a more stable annotation signal.

\subsection{Reproducibility}
\label{sec:reproducibility}
Ensuring reproducibility in LLM-based annotation workflows presents substantial challenges. Specifically, LLM-assisted workflows depend on multiple interacting components that may change over time, including prompts, model versions, provider-side implementations, hyperparameter settings, preprocessing pipelines, and deployment environments. Even minor modifications to prompts or inference settings can lead to substantially different annotation outcomes. In addition, many commercial API providers do not fully document backend model updates or infrastructure changes, making long-term replication difficult.

Improving reproducibility therefore requires both thorough documentation of the annotation workflow and careful consideration of the computational environment in which models are deployed. Researchers should systematically record prompts, model configurations, and annotation outputs, while also considering whether local deployment of open-weights models may be preferable for projects requiring stronger reproducibility guarantees, greater transparency, or stricter data governance.

\paragraph{Metadata documentation.}
Reproducibility requires detailed documentation of all components involved in the annotation process. This includes prompt templates, model identifiers and version numbers (preferably linked to model cards), data descriptions (e.g., via data cards), hyperparameter settings, and timestamps of API calls. Maintaining such metadata facilitates both transparent reporting and future replication attempts.

In practice, documenting only the final prompt template is often insufficient. Prompt engineering is typically an iterative process, and even small prompt revisions may substantially alter model behavior. Researchers should therefore maintain detailed logs of prompt requests, model configurations, and model outputs throughout the annotation workflow. Such logs provide an audit trail of how annotations were generated and allow researchers to inspect, reproduce, or debug individual model decisions. 

The notebook accompanying this paper demonstrates how to automatically generate structured logs in both CSV and JSONL formats containing prompt texts, request identifiers, model providers, model names, inference settings (e.g., temperature, token limits, random seeds), generated scores, and reasoning outputs. An example implementation is shown below:

\vspace{1em}
\noindent\textbf{Python Code} (Appendix~\ref{r_reproducibility} for R code):

\begin{minted}{Python}
# Build a prompt + decision log
os.makedirs("logs", exist_ok=True)
timestamp = datetime.now().strftime("%Y%m%d_%H%M%S")
log_csv_path = f"logs/prompt_log_{timestamp}.csv"
log_jsonl_path = f"logs/prompt_log_{timestamp}.jsonl"

# Score multiple conversations with structured output
multiple_structured_responses = {}
prompt_logs = []

for id, conversation in tqdm(
        zip(test_ids, test_conversations),
        total=len(test_ids),
        desc="Processing Messages"):

    prompt_request = prompt_template.invoke(
        {"conversation": conversation}
    )
    structured_model = model.with_structured_output(
        SpecificityFormat
    )
    structured_response = structured_model.invoke(
        prompt_request
    )
    multiple_structured_responses[id] = structured_response

    # Log prompt, configuration, and output
    messages = prompt_request.to_messages()
    prompt_text = "\n\n".join([
        f"{m.type.upper()}: {m.content}"
        for m in messages
    ])
    prompt_logs.append({
        "chat_id": id,
        "provider": "HuggingFace",
        "model": model.llm.model,
        "temperature": model.llm.temperature,
        "max_tokens": model.llm.max_new_tokens,
        "seed": model.llm.seed,
        "prompt_text": prompt_text,
        "score": structured_response.goal_specificity,
        "reasoning": structured_response.reasoning,
    })

# Save logs to CSV and JSONL
prompt_log_df = pd.DataFrame(prompt_logs)
prompt_log_df.to_csv(log_csv_path, index=False)
prompt_log_df.to_json(
    log_jsonl_path,
    orient="records",
    lines=True
)
\end{minted}

Such logging practices support reproducibility, quality control, and downstream analyses of prompt sensitivity or model drift across providers and model versions. They can also assist reviewers and future researchers in understanding how annotations were generated. However, prompt logs may contain sensitive information, copyrighted material, or personally identifiable data. Researchers should therefore carefully review, anonymize, or partially redact logs before public release.

\paragraph{Version control and archiving.}
To further support reproducibility, annotation scripts, prompts, logs, and metadata should be maintained in a version-controlled repository (e.g., GitHub or GitLab). For long-term preservation and citation, researchers may additionally archive tagged releases on platforms such as Zenodo or OSF. Combining detailed prompt logs with version-controlled code and archived releases helps ensure that future researchers can more accurately reconstruct annotation workflows, even if model versions, provider implementations, or API defaults evolve over time.

\paragraph{Local deployment.}
Researchers who prefer not to rely on external cloud APIs may instead deploy open-weights models locally. Local deployment can improve reproducibility by reducing dependency on external providers and enabling researchers to preserve exact model versions, quantization settings, and runtime configurations. Local deployment may also be desirable when working with sensitive or restricted data that cannot be transmitted to third-party services.

One accessible approach is to use \texttt{Ollama}, which provides a lightweight interface for downloading and serving open-weights language models locally. After installing Ollama\footnote{See instructions at \url{https://ollama.com/}}, researchers can download a model and expose it through a local API endpoint that behaves similarly to cloud LLM APIs. For example, the following terminal commands download and start a small test model:

\begin{minted}{Bash}
ollama pull smollm:135m
ollama serve
\end{minted}

By default, Ollama exposes a local endpoint at \url{lhttp://localhost:11434}, which can then be connected to existing LLM workflows and orchestration libraries such as \texttt{LangChain} and \texttt{ellmer}. The example below demonstrates how a local model can be initialized with \texttt{LangChain}:

\vspace{1em}
\noindent\textbf{Python Code} (Appendix~\ref{r_ollama} for R code):
\begin{minted}{Python}
# Use a local model via Ollama
local_model = init_chat_model(
    "smollm:135m",
    model_provider="ollama",
    base_url="http://localhost:11434",
    temperature=0,
    max_tokens=1000,
    seed=123,
)

# Try a single local call with a single prompt request defined earlier
local_response = local_model.invoke(prompt_request)
print(local_response.content)
\end{minted}

The example above uses the extremely small smollm:135m model purely for demonstration and testing purposes. In practice, researchers will typically require substantially larger models to achieve useful annotation quality. Ollama supports a wide range of openly available models, including instruction-tuned variants from the Qwen, Llama, Gemma, Mistral, and DeepSeek families.

Despite the advantages of local deployment, this approach also introduces practical trade-offs. Running larger models may require substantial computational resources, including high-memory GPUs, large amounts of disk space, and technical expertise related to model serving and hardware configuration. Furthermore, inference speed on consumer CPUs can be considerably slower than cloud-hosted APIs. Consequently, cloud-based APIs may remain the most accessible entry point for many SSH researchers, while local deployments provide an important alternative for projects prioritizing reproducibility, privacy, institutional governance requirements, or long-term archival stability.

\subsection{Common Pitfalls}

\paragraph{Data contamination.}
Avoid publishing or sharing your data and labels before completing the annotation experiment. Public exposure of the dataset may lead to unintended inclusion of your materials in the model's training corpus, thereby inflating performance.

\paragraph{Data leakage in few-shot examples.}
When providing few-shot examples within the prompt, ensure that these examples are drawn exclusively from the development portion of your data. Leakage of test instances or related materials into few-shot demonstrations can yield overly optimistic performance estimates and compromise validity.

\subsection{Summary}
This section highlights practical strategies for improving annotation efficiency, performance, and reproducibility. Caching and batch processing can substantially reduce computational costs, while careful hyperparameter tuning and consistency prompting can improve annotation quality. Transparent documentation and version control help ensure that annotations remain verifiable and reusable. Local deployment provides greater control, including stronger reproducibility and adherence to governance policies. Finally, avoiding contamination and leakage safeguards the integrity of both model evaluation and scientific inference.

\section{Conclusion and Outlook}
This paper presented a practical and methodological guide for using LLMs in text annotation workflows for SSH research. Across the full pipeline (i.e., from project setup and prompt development to API-based prompting, evaluation, reproducibility practices, and downstream statistical analysis), the central message has been consistent: LLMs should be treated as measurement instruments rather than as perfect black-box label generators. Their potential for scaling and accelerating annotation is substantial, but so is the need for explicit quality control, transparent documentation, principled validation, and careful consideration of reproducibility.

A central contribution of this paper is the integration of annotation practice with statistical inference and measurement theory. High agreement scores alone are insufficient if prompt selection is overfitted to development data, if annotation uncertainty is ignored in downstream analyses, or if computational workflows cannot be reproduced. The Explore $\rightarrow$ Select $\rightarrow$ Evaluate framework, the distinction between random and systematic annotation error, and the discussion of downstream correction strategies together provide a coherent path from exploratory prompt engineering to defensible empirical conclusions.

At the same time, reproducibility remains a major challenge for LLM-assisted research. Prompt formulations, provider-side implementations, model versions, and API defaults may all change over time, potentially affecting annotation outcomes even when workflows appear superficially identical. The paper therefore emphasized the importance of systematic metadata documentation, prompt and response logging, version control, and computational archiving practices. We further discussed local deployment of open-weight models as an important alternative for projects prioritizing reproducibility, transparency, privacy, or long-term archival stability.

Looking ahead, several developments are especially important for SSH applications. First, stronger open-weight models and more accessible local deployment infrastructures may reduce dependence on proprietary cloud APIs and improve reproducibility in sensitive research settings. Second, more robust methods for uncertainty quantification and measurement-error correction are needed to better integrate LLM-generated annotations into inferential statistical workflows. Third, the field would benefit from improved tooling for reproducible annotation pipelines, including automatic logging, structured evaluation frameworks, and integration with version-controlled computational environments. Finally, future progress will likely depend not only on advances in model capability, but also on continued integration of insights from psychometrics, causal inference, measurement theory, and reproducible computational social science.

In short, LLM-assisted annotation is most valuable when embedded within a transparent and principled research design: clear construct definitions, iterative but well-documented prompt development, independent evaluation procedures, explicit treatment of annotation uncertainty, and reproducible computational workflows. Under these conditions, LLMs can meaningfully expand the scale, speed, and methodological rigor of text-based SSH research.

\section*{RESOURCE AVAILABILITY}
\subsection*{Lead contact}
Requests for further information and resources should be directed to and will be fulfilled by the lead contact, Qixiang Fang (q.fang@uu.nl).

\subsection*{Materials availability}

Python and R notebooks accompanying this paper can be generally accessed via \url{https://github.com/sodascience/workshop_llm_data_collection} and have been deposited to \url{https://zenodo.org/records/20073016}.

\subsection*{Data and code availability}
All code and example data accompanying this paper can be generally accessed via \url{https://github.com/sodascience/workshop_llm_data_collection} and have been deposited to \url{https://zenodo.org/records/20073016}.

\section*{ACKNOWLEDGMENTS}
This work was funded by the Dutch Research Council (NWO) Large-scale Research Infrastructure Grant, SSHOC-NL (granted under NWO grant number 184.036.020). The authors thank Gabrielle Martins van Jaarsveld for providing the running research example and toy dataset in this paper. AI was used for text proofreading during the writing of this paper.

\section*{AUTHOR CONTRIBUTIONS}

Conceptualization, Q.F., J.G.B. and E.V.K.; 
methodology, Q.F.; 
investigation, Q.F.; 
writing-–original draft, Q.F., J.G.B. and E.V.K.; 
writing-–review \& editing,  Q.F., J.G.B. and E.V.K.; 
funding acquisition, J.G.B. and E.V.K.; 
resources, Q.F., and E.V.K.

\section*{DECLARATION OF INTERESTS}
The authors declare no competing interests.

\section*{DECLARATION OF GENERATIVE AI AND AI-ASSISTED TECHNOLOGIES}
During the preparation of this work, the authors used ChatGPT for writing assistance. After using this tool or service, the authors reviewed and edited the content as needed and take full responsibility for the content of the publication.

\bibliographystyle{numbered}
\bibliography{main}

\appendix
\section{R Code}
\label{r_code}
\subsection{Obtain and Secure API Keys}
\label{r_api_key}
\begin{minted}{R}
# Prompt user for the API key securely (input is hidden) and store it for this session.
groq_api_key <- readline(prompt = "Enter API key for Groq: ")
Sys.setenv(GROQ_API_KEY = groq_api_key)
\end{minted}

\subsection{Minimal Dependencies}
\label{r_dependency}

\begin{minted}{R}
library(ellmer)    # LLM API interfacing in R
library(irr)       # inter-rater reliability (Krippendorff's Alpha)
library(cli)       # progress bars
library(jsonlite)  # JSONL logging
library(tidyverse) # data manipulation and visualization
\end{minted}

\subsection{Data Preparation}
\label{r_load_data}
\begin{minted}{R}
# Build the download URL for the example CSV dataset.
data_url <- paste0(
  "https://sodascience.github.io/workshop_llm_data_collection/data/",
  "srl_data_example.csv"
)
# Load the CSV from the URL into a pandas DataFrame.
df <- read_csv(data_url)
\end{minted}

\begin{minted}{R}
# Use these 10 rows to define test_ids and test_conversations
test_ids <- df |> slice(1:10) |> pull(id) |> as.character()
test_conversations <- df |> slice(1:10) |> pull(conversation)
\end{minted}

\subsection{A Single Prompt Demo}
\label{r_single_prompt}

\begin{minted}{R}
# Build the system prompt
system_prompt <- "
You are an expert in educational assessment and goal evaluation, with
specialized expertise in applying deductive coding schemes to score the quality
and content of student goals.

##TASK##
A university student was given a series of prompts, guiding them through the
process of setting and elaborating on an academic goal for the coming week. You
will be provided with the entire conversation including the prompts, and the
student answers. Your objective is to assess the specificity of of the student's
goal on a scale of 0 to 2 based on the entire conversation.
"
\end{minted}

\begin{minted}{R}
# Define a function that makes a call to the API.
make_chat <- function(system_prompt = NULL) {
  api_args <- list(
    temperature = 0, 
    max_tokens = 1000,
    seed = 123
  )
  chat_groq(
    model         = "meta-llama/llama-4-scout-17b-16e-instruct",
    system_prompt = system_prompt,
    api_args      = api_args,
    echo          = "none"
  )
}
\end{minted}

\begin{minted}{R}
# Create a chat and make the API call to get a single response
chat <- make_chat(system_prompt = system_prompt)
single_response <- chat$chat(test_conversations[[1]])

# Inspect the first response
cat(single_response)
\end{minted}

\subsection{Multiple Prompts Demo}
\label{r_multi_prompts}

\begin{minted}{R}
# Iterate over the conversations and score them
multiple_responses <- list()

cli_progress_bar("Processing Requests", total = length(test_ids))

for (i in seq_along(test_ids)) {
  chat <- make_chat(system_prompt = system_prompt)
  multiple_responses[[test_ids[[i]]]] <- chat$chat(test_conversations[[i]])
  cli_progress_update()
}

cli_progress_done()
\end{minted}

\subsection{Structuring LLM Annotations}
\label{r_structured_output}

Below, we define our desired output format:

\begin{minted}{R}
# Define the expected structured output schema
schema <- list(
  type = "object",
  additionalProperties = FALSE,
  required = c("goal_specificity", "reasoning"),
  properties = list(
    goal_specificity = list(
      type = "integer",
      description = "Score for goal specificity. Only return an integer from 0 to 2",
      minimum = 0,
      maximum = 2
    ),
    reasoning = list(
      type = "string",
      description = "The reasoning to justify the score"
    )
  )
)

output_structure <- type_from_schema(
  jsonlite::toJSON(schema, auto_unbox = TRUE)
)
\end{minted}

To force the LLM to produce outputs in formats specified by you, you need to use the `\$chat\_structured()` method instead of the `\$chat` method.
\begin{minted}{R}
# Score multiple conversations with structured output
multiple_structured_responses <- list()

cli_progress_bar("Processing Messages", total = length(test_ids))

for (i in seq_along(test_ids)) {
  chat <- make_chat(system_prompt = system_prompt)
  multiple_structured_responses[[test_ids[[i]]]] <- chat$chat_structured(
    test_conversations[[i]], type = output_structure
  )
  cli_progress_update()
}

cli_progress_done()
\end{minted}

\subsection{Evaluate Annotation Quality}
\label{r_annotation_quality}
The \texttt{kripp.alpha()} function from the \texttt{irr} package can be used to calculate Krippendorff's~$\alpha$.
We now assess the agreement between the specificity scores generated by the LLM and those coded by human experts.

\begin{minted}{R}
# Compare agreement between expert and LLM ratings
expert_specificity_scores <- df |> slice(1:10) |> pull(expert_specificity_score)
structured_llm_scores <- map_int(multiple_structured_responses, "goal_specificity")

rating_matrix <- rbind(structured_llm_scores, expert_specificity_scores)

cat("Krippendorff's Alpha:")
kripp.alpha(rating_matrix, method = "interval")
\end{minted}

\subsection{Metadata Documentation}
\label{r_reproducibility}
\begin{minted}{R}
# Build a prompt + decision log
dir.create("logs", showWarnings = FALSE)
timestamp      <- format(Sys.time(), "%Y%m%d_%H%M%S")
log_csv_path   <- paste0("logs/prompt_log_", timestamp, ".csv")
log_jsonl_path <- paste0("logs/prompt_log_", timestamp, ".jsonl")

# Score multiple conversations with structured output
multiple_structured_responses <- list()  # Store the structured responses
prompt_logs                   <- list()  # Store the prompt logs

cli_progress_bar("Processing Messages", total = length(test_ids))
for (i in seq_along(test_ids)) {
  chat <- make_chat(system_prompt = system_prompt)
  structured_response <- chat$chat_structured(test_conversations[[i]], 
  type = output_structure)
  multiple_structured_responses[[test_ids[[i]]]] <- structured_response

  # Log the prompt, model config, and output
  prompt_logs[[i]] <- list(
    chat_id       = test_ids[[i]],
    provider      = chat$get_provider()@name,
    model         = chat$get_model(),
    temperature   = chat$get_provider()@extra_args$temperature,
    max_tokens    = chat$get_provider()@extra_args$max_tokens,
    seed          = chat$get_provider()@extra_args$seed,
    system_prompt = chat$get_system_prompt(),
    user_prompt   = test_conversations[[i]],
    output_format = if (is.null(output_structure)) NA else output_structure@json,
    score         = structured_response$goal_specificity,
    reasoning     = structured_response$reasoning
  )
  cli_progress_update()
}
cli_progress_done()

# Save logs to CSV and JSONL
prompt_log_df <- bind_rows(prompt_logs)
write_csv(prompt_log_df, log_csv_path)
# Write each record as a separate JSON line (JSONL format)
write_lines(
  sapply(prompt_logs, \(x) toJSON(x, auto_unbox = TRUE)),
  log_jsonl_path
)

# View the prompt log dataframe
head(prompt_log_df, 3)
\end{minted}

\subsection{Local Deployment with Ollama}
\label{r_ollama}

\begin{minted}{R}
# Use a local model via Ollama
local_chat <- chat_ollama(
  model         = "smollm:135m",
  base_url      = "http://localhost:11434",
  system_prompt = system_prompt,
  api_args      = list(temperature = 0),
  echo          = "none"
)

# Try a single local call with the first conversation defined earlier
single_local_response <- local_chat$chat(test_conversations[[1]])
cat(single_local_response)
\end{minted}

\section{Final Refined Prompt}
\label{final_prompt}
\begin{quote}

You are an expert in educational assessment and goal evaluation, with specialized expertise in applying deductive coding schemes to score the quality and content of student goals. You have a deep understanding of scoring rubrics and are highly skilled at analysing goals for specific characteristics according to well-defined criteria.

\vspace{1em}
\noindent
\#\#TASK\#\#

\noindent
A university student was given a series of prompts, guiding them through the process of setting and elaborating on an academic goal for the coming week. You will be provided with the entire conversation including the prompts, and the student answers. Your objective is to assess all the students answers using a scoring rubric that evaluates four distinct characteristics of the goal on a scale of 0 to 2, representing a characteristic which is not present, partially present, or fully present. The categories you must score are:

\vspace{0.5em}
\noindent
1. Specificity - Goal must be specific rather than general. The context and details of the goal should be explicitly stated and described, and all terms are explained.

Score of 0: Extremely broad, with no details about what this goal entails. States the goal using vague terms without providing any descriptions of what they mean. Or the goal is an abstract concept to improve or work towards, without any explanation of how this could be actionable or concrete. 

Score of 1: States an actionable or concrete goal and offers some descriptions of the terms used. However, there are still some vague terms which are not fully described.

Score of 2: No vague terms which are not described. Clearly states the goal and uses clear descriptions to describe exactly what they want to achieve. OR gives a boundary descriptor which offers context to the other unexplained terms in the goal. 

\vspace{0.5em}
\noindent      
2. Measurability - The goal must be measurable, assessable, documentable, or observable. The outcome should be measurable, and it should also be possible to track progress while working on the goal.

Score of 0: Goal which cannot be measured at all. OR A goal with a conceptual or abstract outcome that cannot be measured directly (i.e. prepared, understand, ready)

Score of 1: An outcome goal, which is measurable as being achieved or not. However, does not allow for measurement of progress while working towards achieving the goal.

Score of 2: Goal with a specific outcome measure which not only allows for the measurement of goal achievement, but also for monitoring of progress as they work towards achieving the goal. 

\vspace{0.5em}
\noindent      
3. Importance - There is an explicit reason for the goal which outlines why this goal is important to achieve. This is put in the context of previous experience or future goals.

- Score of 0: No stated reason for this goal. OR The reason or importance of the goal is externally imposed with no personal motivation (i.e., it is due soon, or deadline is approaching.)

- Score of 1: Importance or reason for this goal is explicitly mentioned, but it is vague with no reference to past experiences or future goals. 

- Score of 2: Importance or reason for this goal is explicitly mentioned, and there is a description of one of the following: 1) Context of the importance of this goal in relation to prior experiences or goals, 2) Context of the importance of this goal in relation to broader study goals or future study plans.

\vspace{0.5em}
\noindent
4. Multi-source Planning - Specific steps are mentioned which are directly related to working towards the goal. Schedule must be included for working on and accomplishing the goal including days or times and not just a numbered list. 

Score of 0: No plan included. OR merely a single activity mentioned. 

Score of 1: INCLUDES ONE of the following: 1) Mentions at least two activities, resources, or strategies (in additional to the original goal) that will contribute to achieving the goal OR 2) Specific schedule outlining days or times to work on the goal. 

Score of 2: INCLUDES BOTH of the following: 1) Mentions at least two activities, learning resources, or strategies (in additional to the original goal) that will contribute to achieving the goal AND 2) Specific schedule outlining days or times to work on these activities or the goal. 

\vspace{1em}
\noindent  
\#\#INSTRUCTIONS\#\#

\vspace{0.5em}
\noindent  
1. Understand the scoring rubric:

   - REVIEW the rubric provided for each category to understand the criteria for scores of 0, 1, and 2.

   - IDENTIFY the key elements that distinguish a low score (0) from a high score (2) in each category.
  
\vspace{0.5em}
\noindent  
2. Analyse the conversation in relation to each category:

- SPECIFICITY: ASSESS the extent to which the goal is specific rather than general. Are context and details of the goal explicitly described, and all terms explained? Is the goal concrete and attainable and not something abstract?

- MEASURABILITY: DETERMINE if goal is measurable, assessable, documentable, or observable. Is the outcome measurable, and is it possible to track progress while working on the goal?

- PERSONAL IMPORTANCE: DETERMINE if there is an explicit reason for the goal which outlines why this goal is important to achieve on the basis of previous experience or in the context of future goals.

- MULTI-SOURCE PLANNING: EXAMINE whether there are specific activities mentioned, and whether these activities directly relate to the goal. Is there a schedule included mentioning days or times of day for working on these activities and accomplishing the goal?

\vspace{0.5em}
\noindent  
3. Assign a score for each category:
   - For each category, ASSIGN a score of 0, 1, or 2 based on the rubric.
   - Use the provided scored examples as a reference to ensure consistency with previous assessments.

\vspace{0.5em}
\noindent  
4. Provide a detailed rationale for each score:
   - EXPLAIN why you assigned each score by directly referencing aspects of the goal that meet or fall short of the rubric criteria.

\vspace{0.5em}
\noindent  
5. Check for consistency:
   - DOUBLE-CHECK that each score aligns with both the rubric criteria and the rationale provided. 
   - MAINTAIN OBJECTIVITY by strictly adhering to the rubric without introducing personal biases.

\vspace{1em}
\noindent  
\#\#EDGE CASE HANDLING\#\#

- If a goal is ambiguous or unclear, SCORE it on the lower end.

- If a goal appears to partially meet the criteria for two different scores, SELECT the score that best reflects the majority of the goals characteristics for that category.

\vspace{1em}
\noindent  
\#\#WHAT NOT TO DO\#\#

- Never apply personal opinion or assumptions outside the rubric criteria.

- never give a score without a detailed explanation, even if the scoring seems obvious.

- never modify or assume student intent score the goal exactly as written.

- never ignore the rubric or provided examples when scoring

\vspace{1em}
\noindent  
\#\#EXAMPLE SCORING\#\#

\vspace{0.5em}
\noindent  
Example 1:

[example conversation mentioned here – removed for data privacy reasons]

Example 1 Scoring:

Specificity: Score (Reason)

Measurability: Score (Reason)

Importance: Score (Reason)

Multi-Source Planning: Score (Reason)

\vspace{0.5em}
\noindent  
Example 2: 

[example conversation mentioned here – removed for data privacy reasons]

Example 2 Scoring:

Specificity: Score (Reason)

Measurability: Score (Reason)

Importance: Score (Reason)

Multi-Source Planning: Score (Reason)

\end{quote}

\end{document}